# Optimization of Energy Consumption in Delay-Tolerant Networks

Junran Wang
School of Computer Science
The University of Nottingham
psxjw16@Nottingham.ac.uk

Milena Radenkovic
School of Computer Science
The University of Nottingham
milena.radenkovic@nottingham.ac.uk

**Abstract**: Delay tolerant network is a network architecture and protocol suite specifically designed to handle challenging communications environments, such as deep space communications, disaster response, and remote area communications. Although DTN [1]can provide efficient and reliable data transmission in environments with high latency, unstable connections, and high bit error rates, its energy consumption optimization problem is still a challenge, especially in scenarios with limited resources.To solve this problem, this study combines the Epidemic[2] and MaxProp[3] routing protocols with Machine Learning Models to optimize the energy consumption of DTNs. Hundreds of simulations were conducted in the ONE simulator, and an external real-world dataset from San Francisco taxi mobility traces [54] was imported. Random Forest[4] and Gradient Boosting Machine (GBM)[5] models were employed for data analysis. Through optimization involving Hyperparameter Tuning and Feature Selection, the Random Forest model achieved an R-squared value of 0.53, while the GBM model achieved an R-squared value of 0.65.

We explore how to use these models to predict the best combination of ONE simulator input parameters, including btInterface.transmitSpeed,btInterface.transmitRange,Group.bufferSize,Group.waitTime,Group.router,Group.msgTtl, Events1.interval, and Events1.size, in order to improve delivery probability while reducing overhead ratio. Our multi-dimensional experiment results show that the delivery probability predicted by the random forest model is 0.59 and the overhead ratio is 3.95, while the delivery probability predicted by the GBM model is 0.66 and the overhead ratio is 4.69. Compared with the original simulation results of the ONE simulator, both models successfully improved delivery probability while reducing overhead ratio, thereby optimizing DTN energy consumption.

**Keywords**: Delay Tolerant Network, Delay Tolerant Routing, Energy optimization, Epidemic, MaxProp, Machine Learning, Random Forest, Gradient Boosting Machine, Delivery Probability, Overhead Ratio, Grid Search, Feature Selection.

# I. INTRODUCTION

## 1.1 Background

Delay Tolerant Networks (DTNs)[6] are specialized network architectures designed to handle extreme environments with long delays, unstable connections, and high bit error rates, commonly found in scenarios such as deep space communications, disaster recovery, and remote areas[7]. DTNs provide interoperable communication between networks with poor performance characteristics by operating above the transport layer, offering services such as data storage, retransmission, authenticated forwarding[8], and routing tolerance for network segmentation[9].Energy management is crucial in DTNs due to the need for nodes to retain and transmit data despite communication delays and breaks[10]. Effective energy management impacts network performance and sustainability, especially in remote or disaster-stricken areas. However, optimizing energy use in DTNs is challenging, requiring new methods and algorithms beyond traditional energy management techniques[11].In general, DTNs occupy an important position in the field of modern communication, and the optimization of energy consumption is one of the key challenges. With the emergence of more and more unconventional communication scenarios, such as UAV communication, underwater communication, etc., DTNs and their energy optimization issues are extremely important.

## 1.2 Theoretical underpinning

Energy optimization in DTNs is complex due to their unique characteristics[12]. Traditional algorithms like linear programming and genetic algorithms may not fully apply [13]. Recent advancements in distributed real-time demand monitoring and automatic resource provision, as discussed in [14] offer potential solutions for DTNs. The paper's utilization of distributed predictive analytics and deep reinforcement learning could inform adaptive energy management in DTNs. [15] introduces a fully distributed energy-aware opportunistic charging approach. This strategy could be adapted to DTNs to address the unique energy management challenges posed by unstable and disconnected networks. Additionally, [16] proposes an adaptive approach to energy management that could be particularly beneficial in the context of DTNs. CognitiveCharge's[16] use of implicit predictive hybrid contact, and resources congestion heuristics could be integrated into DTN energy optimization strategies. In order to save energy without sacrificing data quality during transmission, new routing algorithms and caching strategies may need to be developed. Based on the above considerations, this study adopts two protocols, Epidemic and MaxProp, as solutions for energy optimization in DTN. These protocols can maintain communication functions in the presence of unstable network connections or network partitions and have been shown to be effective in increasing transmission efficiency by widely disseminating information or prioritizing the delivery of information that has not yet been delivered.

## 1.3 Aims & Objectives

The core goal of this research is to optimize energy consumption in DTNs while balancing energy efficiency and network performance. This involves protocol selection and analysis, actual experiment simulation, and simulation parameter optimization.

### 1.3.1 Protocols selection and analysis

Two protocols commonly used in DTN are selected: Epidemic and MaxProp. The Epidemic protocol performs well in environments with unstable connections due to its simple diffusion mechanism, while MaxProp optimizes routing by dynamically calculating the delivery probability. These two protocols have been shown to be effective in exploring energy consumption in DTNs in past studies. Through in-depth analysis of the performance of these two protocols in specific scenarios, we can understand how to use these two protocols to achieve more efficient energy utilization in the case of unstable connections and network segmentation.

### 1.3.2 Actual experimental simulation

Through the *ExternalMovement* function of the one simulator, the San Francisco taxi data set is imported externally. The San Francisco taxi data set contains complex and changeable movement patterns, so it is very suitable for simulating real scenes in DTN. Using this real data, we can more accurately evaluate the performance of Epidemic and MaxProp protocols in real environments. In addition, this practical simulation method ensures the authenticity and reliability of the experiment.

### 1.3.3 Simulation parameter optimization

Run the ONE simulator to conduct simulation experiments to collect data sets; use *btInterface.transmitSpeed*, *btInterface.transmitRange*, *Group.bufferSize*, *Group.waitTime*, *Group.router*, *Group.msgTtl*, *Events1.interval*, *Events1.size* as input variables, and *delivery_prob* and *overhead_ratio* as the target variables. Mining the best input variable combinations from simulation results through machine learning models such as random forest and GBM The main goal of this step is to increase *delivery_prob* while reducing *overhead_ratio*.

This study analyzes the actual performance of the Epidemic and MaxProp protocols on the San Francisco taxi dataset, providing an empirical basis for the application of these two protocols in the specific environment of San Francisco. Through the random forest and GBM models, the parameter combination that improves the network performance in a specific scenario is found, which provides a valuable reference for parameter selection in similar urban traffic. More importantly, this study proposes a specific energy optimization strategy that reduces the overhead ratio while increasing the delivery probability. At the same time, this study also reveals the specific steps for importing external data into the ONE simulator.

In the context of tracking movements of taxis in San Francisco, Bluetooth is more suitable for handling the common short-range and intermittent connections found in urban environments compared to Wi-Fi[17]. Additionally, Bluetooth's low energy consumption makes it particularly practical in energy-constrained mobile environments. Therefore, for the DTN simulation, *btInterface* was intentionally chosen over WiFi.

## 1.4 Discussion

Although Delay Tolerant Network have received extensive attention and research, specific solutions on how to effectively reduce energy consumption are still relatively insufficient. Current research mainly focuses on the theoretical level, exploring various new routing protocols and algorithms. This process is not only time-consuming and resource-intensive but also extremely challenging, as energy consumption optimization requires comprehensive consideration of many complex factors, such as network topology, node mobility, transmission distance, and device energy consumption. Such a theoretical focus has led to the serious neglect of research on the practical application of DTN. There is insufficient evidence to support the feasibility and effectiveness of these theories in the real world, thus limiting the translation of theoretical results into practical applications.

To bridge this gap, we use San Francisco taxi mobility traces combined with Epidemic and MaxProp protocols. We use 40 vehicles in one day to provide empirical support for DTN energy optimization. Simultaneously, we employ random forest and GBM models to seek the optimal combination of input variables for the ONE simulator in the context of the San Francisco transportation scenario. The aim is to effectively reduce energy consumption while ensuring network performance. Additionally, my research reveals specific steps on how to import external data into the ONE simulator. This process provides guidance for simulation and testing in other real-world scenarios. [14] employs distributed predictive analytics and deep reinforcement learning to perform real-time demand monitoring and automatic resource allocation under dynamically changing workload patterns. This offers valuable guidance for addressing the challenges of unstable and disconnected network connections in our research. Similarly, [16] proposes a fully distributed energy-aware opportunistic charging method that adapts to complex network environments through multi-layer real-time multi-dimensional predictive analysis. This provides insights into how we can more accurately capture, predict, and adapt to dynamic space-time energy supply and demand in DTNs.

My research may have some potential value, but it also reveals some practical limitations. First, although the energy consumption analysis based on the 2008 San

Francisco taxi dataset may provide a reference for network performance improvement in traffic scenarios in the same area, However, whether it can be applied to other cities or more complex transportation systems still needs further exploration and verification. Secondly, although the specific steps for how to import external data into the ONE simulator are helpful for simulation and testing in other practical scenarios, this method may need to be adjusted for different data structures. Pre-processing methods during import may vary depending on the specific properties of the data and research needs, so more flexible and customized processing may be required.

## 1.5 Plan

The core goal of this research is to explore the optimization of energy consumption in Delay Tolerant Networks through practical simulations. The specific method is as follows: Firstly, select the trajectories of 40 taxis on May 21 from the San Francisco taxi data set in 2008 and perform data preprocessing to meet the format requirements of the ONE simulator. Subsequently, by adjusting the values of parameters such as *btInterface.transmitSpeed*, *btInterface.transmitRange*, *Group.bufferSize*, *Group.waitTime*, *Group.router*, *Group.msgTtl*, *Events1.interval*, *Events1.size* and other parameters, hundreds of simulation experiments Record the simulation results of *delivery_prob* and *overhead_ratio* and integrate these data into a data set. Then, apply the Random Forest and GBM Models and focus on finding the optimal combination of *delivery_prob* and *overhead_ratio* parameters in the specific traffic environment of San Francisco in order to effectively reduce the energy consumption of the network while ensuring the reliability of message delivery.

The article follows the structure outlined below. The Literature Review section provides a chronological overview of the historical development of energy optimization in Delay Tolerant Network, from the early stages to the latest research trends. This section also identifies gaps in existing research and discusses strategies to address these gaps. Subsequently, the Methodology section introduces the San Francisco taxi dataset, the ONE simulator, and outlines the project implementation process. This process encompasses preprocessing the San Francisco data, importing external datasets into the ONE simulator, constructing, and optimizing machine learning models, and determining the optimal input variable prediction method for the ONE simulator. The Results section presents a comparison of the performance of the Epidemic and MaxProp routing protocols, a parameter sensitivity analysis, and the predicted results of the optimal input variable combination for the ONE simulator. In the Discussion section, the experimental results are comprehensively analysed and contrasted with prior research in the field of DTN energy optimization. Finally, the Conclusion section summarizes the key findings of the study, acknowledges the limitations of the research, and outlines potential directions for future studies.

# II. LITERATURE REVIEW

## 2.1 Introduction

DTN is getting a lot of attention today in networking technologies, especially in modern mobile and IoT applications, which often involve intermittent or delayed connections[18]. As these networks grow in complexity and scope, so do concerns about their energy efficiency. The central focus of this literature review is the problem of energy consumption optimization in DTN. As just mentioned, with the widespread application of mobile devices and wireless sensor networks, reducing energy consumption while ensuring performance is an important challenge. This paper aims to provide an in-depth study of various optimization methods in order to improve the energy efficiency of DTNs.

The research scope of this literature review will include the following topics: an overview of Delay Tolerant Network; energy management in DTN transmission protocols; how to ensure reliable data transmission under intermittent connections and limited resources; clustering and analysis in DTN applications; Practical Applications of Markov, Reinforcement Learning, and Bayesian Algorithms to DTNs. This literature review will not address the following topics: energy management in wired networks or wireless networks with stable connections; routing protocols for wired networks; machine learning techniques for visual recognition or language processing; fixed network topologies; and energy efficiency Irrelevant Network Optimization Methods

This literature review firstly introduces the theoretical basis of DTN and energy consumption. Then, it will be divided into three parts in chronological order: early optimization strategy, which discusses the initial stage of DTN energy optimization; mid-term optimization strategy, which analyzes the key progress in this field in the mid-term stage; and near-term optimization strategy, which describes the recent development trend and status condition. Finally, in the overall section, the existing research will be summarized, and some possible research gaps and future directions will be briefly pointed out.

## 2.2 Theoretical Basis: DTN and Energy Consumption Optimization

### 2.2.1 Basic concepts and principles of DTN

Delay Tolerant Network is a unique network architecture designed to adapt to different network environments with poor performance characteristics, especially when the network connection is unstable or discontinuous[19]. DTN operations are based on the abstract concept of message exchange, where messages are aggregated into "bundles" and processed by specific devices, such as "bundle forwarders" or DTN gateways. When reliable delivery of messages needs to be guaranteed, DTN gateways will send messages Stored in memory in a non-volatile manner and map information between different transmissions through name resolution[20]. DTN gateways also perform authentication and access control on incoming traffic to ensure that they are allowed to forward; DTN architecture consists of regions and is composed of DTN gateways, in which regional DTN gateways are connected to each other and are responsible for the cross-regional forwarding of data packets, which is different from the working mechanism of ARPANET gateways[8]; In unstable or discontinuous connection environments, DTN moves messages by using specific time-related connections. Some application scenarios for delay tolerant networks include: space networks, where communication links are intermittent and have long propagation delays; military networks, where communication links are unreliable and have high latency; disaster response networks, where communication infrastructure may be damaged or destroyed; remote and rural networks, where communication infrastructure is limited or nonexistent; and vehicular networks, where communication links are intermittent and have high latency[21]. Challenges faced by DTNs include: bundle routing, which presents difficulties in routing bundles through the network due to the intermittent and unpredictability of communication links; bundle fragmentation and reassembly, which requires fragmentation and reassembly of bundles in DTNs to accommodate the limited resources of end nodes package; Security:DTN needs to safeguard communication links and prevent unauthorized access to the network. interoperability, DTN needs to achieve interoperability between different networks with diverse performance characteristics and protocols; Naming and Addressing: DTN needs to provide globally unique naming and addressing schemes while accommodating the limited resources of end-nodes[22].

### 2.2.2 The Importance and Challenges of Energy Consumption

Energy efficiency improves network longevity. [23] proposed the Message Ferrying (MF) method. This approach utilizes a special set of mobile nodes, called message ferries, that provide communication services to nodes within the deployment area. The MF design improves data delivery performance and reduces the energy consumption of nodes by exploiting mobility. By reducing energy consumption, nodes can run for longer periods of time, extending the lifetime of the entire network. Energy efficiency can promote sustainable development. [24]highlighted the role of energy harvesting technologies in promoting sustainable development. By capturing and storing energy from renewable sources, energy harvesting technology not only reduces reliance on traditional power sources, thereby enhancing the overall energy efficiency of the system, but also minimizes the impact of energy consumption on the environment. This helps in reducing carbon emissions, improving resource management, and enhancing energy security, thereby contributing to sustainable development. Energy efficiency plays an important role in disaster response networks. Reasonable energy management can ensure the continuity and stability of rescue communications, which is crucial for post-disaster rescue operations. [10] proposed an energy efficient DTN framework that helps provide different disaster management services during such operations. The simulation results show that the three-layer architecture combined with the MaxProp routing algorithm exhibits excellent delivery probability and average energy consumption per message delivery, making it an effective choice for post-disaster communication.

Finding the balance that maintains energy efficiency without compromising critical performance such as latency and reliability is a challenge. [23] pointed out that traditional data transmission strategies could lead to inefficiencies and increased delays, thereby compromising these crucial performance aspects. Additionally, overly focusing on energy efficiency optimizations might result in longer delays or decreased reliability. Conversely, pursuing critical performance aspects to the extreme could lead to increased energy consumption. Therefore, in sparse mobile ad hoc networks, balancing energy efficiency and critical performance to achieve optimal performance is a complex issue that requires careful consideration of technical details and precise parameter adjustments. Another issue is the trade-off between security and energy efficiency. Typically, ensuring network security demands more computational and transmission resources, which may conflict with the goal of energy efficiency. Achieving a solution that is both secure and energy-efficient requires careful design and trade-offs. For instance, [25] proposed a peer-to-peer indexing structure called Peer-to-Peer Tree, aiming to achieve minimal energy consumption and shortest query processing response time, thereby balancing security and energy efficiency. Additionally, [10] point out that most current DTN architectures place little emphasis on energy efficiency. They have observed that, so far, research has not addressed the impact of different mobility patterns on existing DTN routing algorithms in terms of data propagation and energy consumption. Existing DTN architectures primarily consist of three major types of devices, and the energy consumption of these devices varies significantly. At the same time, although near-line-of-sight Wi-Fi towers can serve as fourth-layer devices to provide routing services and may be beneficial in areas physically isolated or disconnected due to disasters, their use in DTN architectures is rare.

### 2.2.3 Basic Energy Optimization Methods and Strategies

After the above-mentioned DTN background and the importance and challenges of energy consumption, it is particularly critical to understand and discuss energy optimization methods and strategies applicable to these networks. Due to the unique nature of DTN, especially its intermittent and unstable connection characteristics, flexible and targeted strategies for energy management are required.

In terms of energy management for DTN, there are generally three strategies to optimize energy efficiency. Firstly, the scheduling strategy focuses on how to reasonably arrange the transmission and processing of messages under different load and connection modes[26]; secondly, the transmission control focuses on dynamically adjusting the transmission rate or transmission power according to the network status and node energy conditions, thereby reducing energy consumption[27]; Finally, power management discusses how to adjust the node's working and sleeping modes and choose the appropriate hardware and software configuration without sacrificing critical performance[28].

Among the many routing and energy optimization protocols for DTN, this study focuses on two protocols, Epidemic and MaxProp, because they may be more advantageous in specific application scenarios such as disaster response and mobile networks. These two protocols were chosen as research objects based on their performance in practical applications, and both are closely related to the specific urban traffic scenarios and goals that this research focuses on[29]. Epidemic protocol: Because of its simplicity, robustness, and scalability, it is widely used in environments with unstable and intermittent connections. Its working principles, strengths, and limitations make it ideal for studying energy optimization in a variety of scenarios[30]. The MaxProp protocol, compared with the Epidemic protocol, provides a more complex and fine-grained routing strategy that can better balance energy consumption and performance under certain conditions. Its importance in DTN energy optimization has been demonstrated, especially when flexible routing and limited resources are required[31].

The Epidemic protocol is generally recommended for connecting intermittent and unpredictable sparse networks. The core idea of epidemic routing is similar to the spread of viruses. Packets are assigned a special ID linked to them and all their replicas, and they are forwarded using a first-come-first-served (FCFS) order. During a contact between two nodes, packet transmission will continue until the contact period ends or all uncommon packets are exchanged, resulting in the same list of packets between the two nodes. Given sufficient ties, buffer sizes, and packet lifetimes, epidemic routing ensures that each packet reaches its destination, possibly taking the shortest route[32]. The Epidemic protocol provides a powerful solution in sparse or highly dynamic DTN environments. Its strengths lie in ensuring message delivery and minimizing end-to-end latency by utilizing all possible routing paths. However, the disadvantage of this approach is its huge consumption of limited resources such as memory, energy, and connection duration. This can lead to inefficiencies and higher costs in energy-constrained scenarios. Overall, effective deployment of the Epidemic protocol requires carefully balancing its delivery and latency benefits with the challenges of resource consumption, especially in energy-sensitive applications[33].

MaxProp is a routing protocol for Delay Tolerant Network, specifically designed to improve message delivery probability, overhead ratio, and median latency in sparsely and intermittently connected networks. It does this by prioritizing packets based on their hop count and delivery likelihood and dividing the buffer into two parts so that packets are prioritized on a drop schedule[34]. The advantage of MaxProp lies in improved message delivery and energy optimization in DTN, which can achieve the same network resource consumption as other routing protocols by reducing overhead ratio and latency. However, this protocol may not be suitable for all types of DTN; performance may vary depending on network topology and conditions; and it may require more computing resources than other routing protocols. Overall, MaxProp may be an effective routing option in specific scenarios through its unique prioritization and energy optimization mechanisms[3].

## 2.3 The changing process of DTN energy optimization

The change process of DTN energy optimization is divided into three stages: Early optimization strategy, Mid-term optimization strategy, and Recent optimization strategy.

### 2.3.1 Early optimization strategy

Early energy optimization strategies mainly focused on network architecture design and Energy Sensing Technology. Research at this stage mainly revolves around how to build an architecture that adapts to limited resources and challenging network environments, and how to extend network life and maintain connectivity by intelligently selecting paths and balancing energy consumption.

2.3.1.1 Network Architecture Design
[8] proposed a new network architecture design for challenging network environments with limited power or memory resources. This architecture addresses routing issues for low-duty cycle operation and aims to optimize energy consumption in challenging networks. To support this design, they implemented key services on top of the network's transport layer, such as in-network data storage, retransmission, interoperable naming, authenticated forwarding, and coarse-grained services. On the other hand, [35] focused on the computational problem of energy efficiency for wildlife tracking via the ZebraNet system. They designed a custom tracking collar system that operates as a peer-to-peer network to achieve a very high "data attribution" success rate. This network architecture is specially designed for animal research in wild areas, fully considering the minimum consumption of energy, storage, and other resources.

When comparing these two strategies, [8]'s scheme focuses on providing a general solution for complex and challenging network environments. Their design philosophy involves constructing a comprehensive and highly adaptable network architecture. On the other hand, [35] place greater emphasis on specific application scenarios, with their design strategy prioritizing efficiency and minimal resource consumption. Although the application scenarios behind it are different from those proposed by Fall et al. The construction concept of its network architecture is also to adapt to the challenge of limited resources.

Although [8]'s network architecture design emphasizes the optimization of general services in challenging network environments and provides a comprehensive solution, it may not be flexible and efficient in some specific application scenarios. For example, in field environments that require extreme energy efficiency, the architecture may be too complex and resource intensive. In contrast, although Juang et al.'s scheme is carefully designed for specific application scenarios and can minimize resource consumption, it is too dependent on specific application backgrounds and requirements and may lack versatility. This may limit its application potential in other types of network environments.

2.3.1.2 Energy Sensing Technology
Shah and Rabaey.[36] proposed an energy-aware routing design for low-energy self-organizing sensor networks. They emphasize that energy-aware routing helps increase the lifetime and long-term connectivity of the network by occasionally using suboptimal paths and ensuring that nodes consume energy in a uniform manner. Compared with other schemes such as directed diffusion routing, their simulation results show that this special design can improve the network lifetime by up to 40%. Schurgers and Srivastava.[37] focused on energy-efficient routing design for wireless sensor networks. Aiming at the problem that traditional routing protocols ignore limited energy, they proposed a new routing technology based on energy histograms. Through two specialized algorithms, including minimizing energy consumption and exploiting energy potential with multi-hop communication, they successfully extended the lifetime of the sensor network. Shah and Rabaey's proposal focus on low-energy self-organizing sensor networks, a particular design that, while efficient for the target environment, may have limited applicability in other types of network environments. In contrast, Schurgers and Srivastava's approach is designed for wider applications of wireless sensor networks, which enhances its generality but may also introduce more complex customization requirements to achieve the best results in specific scenarios.

Inspiration from Early energy optimization strategies: In earlier studies, Shah and Rabaey[36] proposed a design for energy-aware routing that emphasized how to extend network lifetime and maintain connectivity by balancing different factors. This idea played a significant inspirational role in my research on energy optimization in Delay Tolerant Networks. I realized that in energy optimization, the focus should not be solely on one metric but rather on finding the optimal balance between multiple key factors. Therefore, in my approach, I sought to find a balance that would both improve the delivery probability (the chance that a message is successfully delivered to its destination) while also reducing the overhead ratio (the additional communication load generated within the network to achieve effective delivery). These two indicators together reflect the performance and energy efficiency of the network, and their balance provides an effective path for optimizing energy consumption.

**2.3.2 Mid-term optimization strategy**

The research in the mid-term stage of DTN energy optimization mainly focuses on the design of routing protocols and energy-saving mechanisms for energy efficiency. Routing protocols attempt to reduce energy consumption while maintaining network performance, while energy-saving mechanisms aim to reduce ineffective and excessive communications through careful management of communication connections, intelligent caching strategies, and energy management of sensors and mobile devices, thereby extending the time between devices and extending the life of the network. This also ensures the sustainable operation of DTN in an energy-limited environment.

2.3.2.1 Design of Rouing Protocols for Energy
Spyropoulos et al.[38] proposed the "Spray and Wait" routing scheme, which is tailored for DTNs with the aim of optimizing energy usage. This scheme involves "spraying" multiple copies of messages into the network and then "waiting" for one of these copies to reach the destination. The authors conducted experiments using the ONE simulator to demonstrate the effectiveness of this approach in terms of average message delivery delay and the number of transmissions per message. The results showed that the Spray and Wait method outperformed five other routing protocols: Epidemic routing, Randomized flooding, Utility-based routing, Seek and Focus single-copy routing, and Oracle-based Optimal routing. As a highly scalable and easily implementable alternative, the Spray and Wait scheme plays a significant role in optimizing energy consumption for DTNs.

Burgess et al. [3]proposed the MaxProp protocol for vehicle-based Delay Tolerant Networks, with a special focus on how to achieve efficient routing optimization in energy-constrained, challenging environments. By carefully controlling the replication of packets, the protocol not only increases the probability of delivery but also optimizes energy consumption by avoiding excessive replication. In fact, excessive data packet replication may lead to network congestion and energy waste, and MaxProp uses a balancing mechanism to ensure the effective transmission of data while also considering energy efficiency. In order to test and verify the effect of the MaxProp protocol in a real environment, Burgess et al. also deployed hardware for data transmission on 40 buses, each equipped with a Linux operating system, two 802.11 radio modules, GPS, and a 40GB hard drive. Data was transferred between these buses via passing each other and available hotspots.

"Spray and wait" [38]and "MaxProp" [3] are two different DTN energy optimization schemes. "Spray waiting" reduces energy consumption by broadcasting multiple messages and waiting for one of them to arrive, which is suitable for large and dense network environments. In contrast, "MaxProp" is detailed and suitable for vehicle based DTN, which achieves a balance of high delivery rate and energy efficiency by carefully controlling the replication of packets, especially in emergency situations. The former is validated by simulation, and the latter is tested in a real vehicle network. Although both solutions target energy efficiency, their approaches and applicable scenarios are different. "Spray and Wait" is suitable for a wide range of applications, while "MaxProp" pursues precision and efficiency in specific environments.

The "Spray and Wait" scheme [38] does a good job of reducing message delivery latency but may fall short in terms of device energy management and overall energy efficiency. Also, a large number of message copies can cause resource consumption issues. The "MaxProp" protocol of Burgess et al.[3] balances delivery rate and energy consumption through fine-grained control of packet replication but may lack generality in terms of broader application scenarios and may not fully consider other aspects of energy management. Overall, these two groups of researchers have made some important progress in energy optimization in DTN environments, especially in routing design. However, their approach may be too focused on optimization at the routing level and not sufficiently involved in integration with overall energy management strategies.

2.3.2.2 Design of Energy Mechanism
In Delay Tolerant Network, the introduction of throwbox provides a new direction for energy optimization. Throwboxes act as static, battery-powered relay nodes that store and process data, thereby enhancing DTN transmission opportunities and performance, improving network performance, and reducing energy consumption.

Nilanjan Banerjee et al. [39]proposed a throwbox hardware and software architecture design for energy efficiency in DTN. The authors contend that a key aspect of enhancing DTN performance lies in designing more transmission opportunities. However, merely increasing transmission opportunities is insufficient; without effective power management, the benefits of the throwbox can be constrained. Therefore, Nilanjan Banerjee et al. [39]introduced a multi-layered, multi-radio, scalable, solar-powered hardware platform. They employed approximate heuristic techniques to maximize forwarded byte counts. The throwbox employs an approximate heuristic to address an NP-Hard problem under average power constraints while maximizing its forwarded byte counts. Through actual deployment and simulations on the UMassDieselNet testbed, they demonstrated that a single throwbox equipped with a 270-square-centimeter solar panel can operate indefinitely. It elevated data packet delivery rates by 37% in the network and reduced message delivery delays by at least 10%.

On the other hand, Wenrui Zhao et al. [40]proposed a strategy to use throwboxes to enhance the capacity and performance of delay-tolerant networks. Their approach is to design and deploy fixed throwboxes as relays, thereby increasing connection opportunities in the network. By taking routing and placement into account using specific algorithms and employing placement algorithms with limited knowledge about the network structure, they evaluated the effect of throwboxes on improving throughput and latency under node movement rules or using multipath routing. It also analyzes the limitations of using single-path and popular routes and highlights the need for throwboxes in terms of high availability, high data transfer rates, processing power, and energy efficiency.

The studies of Nilanjan Banerjee et al.[39] and Wenrui Zhao et al.[40] both focus on using throwboxes to enhance the performance and energy efficiency of delay-tolerant networks, but the approaches of their studies vary. Banerjee et al. [39]achieved their goals mainly through innovations in hardware, software, and energy management, focused on building an energy-efficient platform, and demonstrated their results through practical tests. Zhao et al. [40]started from the perspective of network structure and algorithm design, studied the optimal deployment and routing strategy of throwboxes, and then improved the energy optimization of DTN.

The approaches proposed by Nilanjan Banerjee et al. [39]and Wenrui Zhao et al.[40] for optimizing DTN energy using throwbox indeed demonstrate the potential to reduce network latency and enhance data throughput. However, they may also pose some challenges. Firstly, energy management is a concern. While Throwbox is designed to optimize energy utilization, balancing the trade-off between energy consumption and network performance is a complex issue. Secondly, network compatibility could also be a challenge. The articles do not mention whether Throwbox is universally applicable to all types of DTN environments or network configurations, indicating that specific routing and mobility patterns might be required.

Inspiration from Mid-term energy optimization strategies：: During the mid-stage of my research, the work of Burgess et al. [3] introduced the MaxProp protocol, which emphasized a balanced mechanism to ensure effective data transmission while considering energy efficiency. This concept led me to consider adopting the MaxProp routing protocol. Additionally, the proposal by Spyropoulos et al. [38] for the Spray and Wait routing protocol highlighted the high transmission rate of Epidemic. Given that I plan to simulate urban traffic scenarios in my experiment, where rapid message propagation is crucial, I have also chosen to incorporate the Epidemic routing protocol.

**2.3.3 Recent optimization strategy**

The Near-term optimization strategy mainly focuses on reducing DTN energy consumption through algorithm design and machine learning methods.

2.3.3.1 Algorithm design method to reduce DTN energy consumption

Jiagao Wu et al.[41] proposed an energy-efficient replication-constrained optimization algorithm based on the Box complex method, which is specifically designed for the energy optimization problem of Delay Tolerant Network. In DTN, nodes are usually energy-limited devices, so efficient routing protocols are urgently needed to improve their performance and extend their service life. The study not only focuses on energy efficiency but also considers the phenomenon of multi-community and social selfishness among people with similar interests in the real world. By using ordinary differential equations (ODEs) to analyze the performance of epidemic routing protocols in multi-community scenarios, their algorithm aims to determine the optimal replication limit in multiple communities to effectively improve energy efficiency. The numerical and simulation results of this study further demonstrate that the proposed routing protocol can effectively reduce energy consumption and analyze the impact of socially selfish behavior.

Bo Yang et al. [42] studied the application of the unmanned aerial vehicle (UAV) ferry algorithm in Delay Tolerant Network. The ferry algorithm for UAV is analyzed from the aspects of UAV network architecture, node routing algorithm, node deployment, node trajectory optimization, node energy optimization, and node storage allocation optimization. Optimize the flight path to reduce energy demand by utilizing the energy minimization UAV trajectory algorithm based on genetic algorithms adopt the deployment algorithm of UAV ferry nodes to find the minimum number of UAVs serving a specific area and their optimal deployment locations to reduce unnecessary energy consumption, and use storage allocation optimization algorithms such as weighted minimum and maximum fairness principles to effectively allocate storage space for ferry nodes, thereby reducing energy consumption and improving network transmission success rate and average network delay.

Both Replication-Restricted Optimization Algorithm by Wu et al. [41] and the Unmanned Aerial Vehicle (UAV) Ferry Algorithm by Yang et al.[42], reflect a thorough exploration of optimizing energy efficiency and network performance. However, their focal points differed. Wu et al. [41] addressed energy optimization by combining social factors and applying a complex approach based on Box constraints. On the other hand, Yang et al. [42]focused on reducing energy consumption and enhancing network performance through the utilization of UAV ferry nodes and employing various methods, such as trajectory optimization based on genetic algorithms.

The studies conducted by Wu et al. [41]and Yang et al. [42]respectively focused on energy optimization issues in Delay Tolerant Network using distinct algorithmic approaches. Wu et al. [41]emphasized the integration of social structures and the enhancement of energy efficiency. Although their theoretical framework was robust, it might be applicable only to networks with well-defined multi-community structures. On the other hand, the drone-ferry algorithm of Yang et al.[42] achieves flexibility and automation through multifaceted optimization but may increase the complexity and cost of the system.

2.3.3.2 Machine Learning Approach Reduces Energy Consumption

Milena Radenkovic et al[16] , proposed CognitiveCharge, an innovative framework designed to optimize vehicle charging in the context of the rapid adoption of electric vehicles (EVs). CognitiveCharge employs implicit prediction of hybrid contacts and resource congestion heuristics to identify nodes and network areas at risk of energy depletion. CognitiveCharge employs implicit prediction of hybrid contacts and resource congestion heuristics to identify nodes and network areas at risk of energy depletion. Additionally, this architecture dynamically reallocates energy from surplus areas to regions facing energy consumption challenges, adapting at a non-uniform pace. Evaluation using multi-day traces from both San Francisco, USA, and Nottingham, UK, demonstrated that CognitiveCharge effectively alleviates congestion at temporary and infrastructure charging points, minimizes vehicle charging wait times, significantly reduces the number of energy-dependent nodes, and thereby lowers DTN energy consumption.

Gechen et al[43]. systematically explored various artificial intelligence frameworks, including feature engineering, classification, detection, and time series prediction, and their applicability in ocean science. Through a series of case studies, the authors demonstrate the efficacy of artificial intelligence methods in solving complex oceanographic challenges such as ocean current reconstruction, Arctic Sea ice prediction, and deep-sea debris detection. This paper highlights the value of integrating ocean physics into artificial intelligence models to improve their learning capabilities. Special attention is paid to utilizing deep learning methods such as convolutional neural networks (CNN) for image analysis and object detection in oceanographic data. To promote the sustainable development of global marine ecosystems and thereby reduce DTN energy consumption in marine scenes.

Milena Radenkovic et al[16]. and Gechen et al[43]. both focus on using machine learning to optimize the energy consumption of DTN. Milena Radenkovic et al[16]. proposed the CognitiveCharge framework, which identifies network nodes and regions that may run out of energy by employing a hybrid heuristic of implicit prediction and resource congestion. However, this approach may face challenges in practical applications due to unpredictable factors such as traffic patterns, weather conditions, and emergencies that may affect its energy supply and demand. On the other hand, Ge Chen et al[43]. applied deep learning techniques, particularly Convolutional Neural Networks (CNNs), in the field of marine science for image analysis and object detection. This approach demonstrates efficacy in tackling complex oceanographic problems. However, due to its computational complexity, this approach may not be suitable in resource-constrained environments such as remote ocean observatories or may require expensive hardware to run.

Inspiration from recent energy optimization strategies: Milena Radenkovic et al[16]. verified the effectiveness of CognitiveCharge by using a real external dataset, which strengthened my idea to conduct experiments using the San Francisco taxi movement trajectory dataset. This kind of real-world data can really increase the practicality of an experiment. Additionally, the work by Ge Chen et al[43]. serves as inspiration for drawing parallels between the sparse and unstable characteristics of marine environments and the taxi network in San Francisco. This comparison leads me to consider employing relevant Machine Learning models, such as Random Forest and GBM models, to mitigate energy consumption.

## 2.4 Research Gap

From early to recent research, the following possible research gaps can be observed: Practical applications of real-world datasets, Specific protocol optimization based on specific environments, and the integration of optimization methods using Machine Learning.

### 2.4.1 Practical applications of real-world datasets

Many studies ignore datasets related to real-world scenarios, such as urban traffic environment aspects. A lot of research focuses on exploring various routing protocols and optimization methods through theoretical analysis and computer simulation, especially DTN energy optimization. While these approaches reveal how protocols and algorithms work and predict DTN energy optimization performance in a theoretical setting, they may ignore real-world dynamics and emergent factors. For example, the San Francisco Taxi dataset includes many unpredictable elements such as traffic jams, traffic lights, and emergencies, reflecting the actual urban mobility in San Francisco. Existing theoretical and simulation approaches may not fully capture these real-world effects due to the lack of studies based on such real-world datasets. Therefore, this shortcoming may lead to insufficient accuracy and applicability of existing methods in real-world settings.

### 2.4.2 Specific protocol optimization based on specific environments

In existing research, for specific protocols such as Epidemic and MaxProp, the comparative analysis and optimization of the two in specific environments may be insufficient. Many studies may tend to adopt a general approach to the optimization of a specific environment, such as energy storage, distribution, and conservation in nodes, which may ignore the unique needs of different protocols manifested in a specific environment[44]. For example, the performance of these protocols may vary significantly when faced with different network topologies, traffic loads, and connection stability. However, the difference in protocol performance in a specific environment may significantly affect the reliability and efficiency of the system in practical applications. For example, an urban transmission system may face complex network topologies, where certain areas may have high traffic loads and the stability of connections may be affected by buildings and moving obstacles. In this case, generic optimization methods may not accurately predict how the protocol will perform in real applications. Therefore, in order to fully exploit the potential of specific protocols, comprehensive comparative analyses and optimizations need to be conducted within specific environments.

**2.4.3 The integration of optimization methods using machine learning**

Although recent DTN energy optimization research has begun to apply machine learning, there may be a research gap in the use of ensemble learning methods to find the optimal parameter combination to improve DTN performance. The ensemble learning method can bring the following benefits.: For specific protocols and environments in DTN, ensemble learning can select the best combination from many parameter combinations to maximize performance indicators such as *delivery_prob*, *overhead_ratio*, etc. This method helps overcome the limitations of traditional methods of parameter adjustment and enables the protocol to run more flexibly and efficiently in specific environments. Ensemble learning methods can more comprehensively consider the impact of different parameters on performance. Especially in the face of complex network topologies and changing traffic loads, it may be difficult for traditional single optimization methods to capture the complex interrelationships among multiple factors. By synthesizing the views of multiple models, ensemble learning can provide more precise results for performance prediction and optimization. Ensemble learning can also enhance the robustness of the optimization process[45]. Given the uncertainty and variability of the environment in DTN, the performance of a single model may fluctuate to a certain extent. By integrating the predictions of multiple models, the instability of a single model can be reduced, resulting in a more reliable performance optimization solution.

**2.5 Approaches to the Research Gap**

For the practical application of real-world datasets. The May 21, 2008, San Francisco taxi dataset was chosen. This dataset not only included daily factors such as traffic congestion and traffic light signals but also reflected the actual urban mobility and unpredictable sudden events in San Francisco. This approach took a step further to ensure the accuracy and applicability of urban traffic simulation in San Francisco, aiding in understanding the impact of real urban flow and traffic issues on energy optimization.

For the Specific protocol optimization based on specific environments. We performed extensive simulations using the San Francisco Taxi Dataset through the ONE simulator. These simulations accurately capture a variety of dynamic factors in a given environment, including changes in traffic flow, the complexity of road layouts and the impact of moving obstacles. I particularly focused on two key evaluation metrics: delivery_prob and overhead_ratio. By enhancing the probability of successful transmissions and reducing the ratio of overhead, I effectively further reduced energy consumption. Throughout the simulation process, to ensure the reliability of the results.

I performed hundreds of simulations runs and adjusted multiple parameters such as btInterface.transmitSpeed, btInterface.transmitRange, etc. This approach enabled me to deeply investigate the specific impact of different parameter combinations on the performance of the Epidemic and MaxProp protocols. Finally, through analysis of the resulting data, I could identify the optimal combinations of these protocols in each environment and apply it to real-world simulations.

For the integration of optimization methods using Machine Learning. In this study, we utilized experimental data generated by the ONE simulator to train Random Forest and Gradient Boosting Machine models. The performance evaluation of the model is based on three key indicators: MSE, RMSE and R-Squared. Following the initial training and evaluation with default parameters, we employed grid search and feature selection techniques for further model optimization. The ultimately optimized models were then used through a differential differentiation method to predict the optimal combination of input variables for the ONE simulator. This aimed to enhance *delivery_prob* while simultaneously reducing overhead_ratio, thereby lowering energy consumption in DTN networks.

**2.6 Overview**

After searching the relevant literature on DTN energy optimization, it was further confirmed that DTN plays a key role in ensuring the continuity and stability of information transmission, especially in scenarios where the network connection is unstable or intermittently disconnected. Although energy consumption is critical to the continued operation of a DTN, it comes with numerous challenges. Despite the existence of some basic routing protocol optimization strategies, they might have difficulty meeting the energy-saving requirements in practical applications. Over time, DTN energy optimization strategies have evolved from early network architecture design and energy-aware technology to mid-term routing protocols and energy-saving mechanisms, and recently to algorithm design and machine learning methods. These studies not only demonstrated progress in the field but also revealed research gaps in areas such as optimization incorporating real datasets, protocol optimization in specific contexts, and ensemble machine learning optimization methods.

# III. METHODOLOGY

**3.1 Introduction**

In the energy optimization of Delay Tolerant Networks, this study selected two specific protocols, Epidemic and MaxProp, and took the San Francisco taxi data set on May 21, 2008, for empirical research. Hundreds of simulations were performed in the ONE simulator, and the optimal parameter combination was predicted by using the random forest and GBM model, Thus, the aim was to enhance the delivery_prob while simultaneously reducing the overhead_ratio. and further improving DTN energy consumption. This section will be divided into protocol selection, San Francisco dataset introduction, ONE simulator setting and simulation process, machine learning model, Introduction to machine learning and modeling optimization process, methodology summary and other subsections to explain this optimization process.

**3.2 Protocol Selection**

As mentioned in the literature review section, the Epidemic and Maxprop protocols were selected for the energy optimization of Delay Tolerant Networks to match the specific traffic scenarios in San Francisco, making the simulation results more authentic and convincing. The Epidemic protocol has been widely used in environments with unstable and intermittent connections due to its simplicity, robustness, and scalability. In a large city like

San Francisco, where taxi connections can be intermittent, the Epidemic protocol is an appropriate option. However, this approach may consume a lot of limited resources, such as memory, energy, and connection time, which may lead to inefficiency in energy-constrained scenarios[33]. Compared with the Epidemic protocol, the MaxProp protocol provides a more complex and detailed routing strategy, which can better balance energy consumption and performance under certain conditions. MaxProp protocol prioritizes data packets, thereby providing more flexible routing in the complex traffic network of San Francisco. However, it may require more computing resources[3]. Overall, the combination of these two protocols covers a broader range of usage scenarios and network conditions in San Francisco, contributing to further improvement in DTN energy consumption.

### 3.3 San Francisco Taxi Dataset

The San Francisco taxi dataset is a representative urban traffic dataset, widely used for studying and analyzing city traffic patterns and flow. This dataset covers a 25-day time frame from May 17, 2008, to June 10, 2008, and comes from the "epfl/mobility dataset" [54]. Each record in the dataset includes the following details: taxi_id, a number that identifies the related taxi; trajectory_id, a numerical value that identifies the trajectory in the original dataset; timestamp, corresponding to the start time of the taxi ride; source_point, representing the origin GPS point of the taxi ride; target_point, representing the destination GPS point of the taxi ride. Coordinates are presented in the EPSG:4326 geodetic coordinate system format, as longitude and latitude. Due to San Francisco's complex road network and unique geographical features, this dataset is able to reflect various city traffic scenarios, from busy commercial areas to tranquil residential zones. This diverse dataset provides assistance in understanding and modeling urban traffic.

### 3.4 One simulator setting and simulation process

#### 3.4.1 Opportunistic Network Environment (ONE) Simulator

Existing mobile ad hoc networks (MANETs) and DTN routing simulators such as ns2, OMNeT++, dtnsim and dtnsim have a series of problems. They lack comprehensive support for DTN networks, which may result in insufficient accuracy when simulating this network type. Second, these simulators have limitations in routing protocols and visualization. At the same time, there is a major problem that they may rely on randomly distributed values, which however can be challenging in depicting the complexities of human behavior in the real world. Additionally, real-world trajectories, while helpful in capturing realistic behavior, may limit the scope of simulator applications due to factors such as low spatial and temporal granularity, fixed populations, and specialization. Therefore, ONE simulator was developed, which has reasonable mobility modeling capabilities, integrated support for DTN routing, and can visualize simulation progress and results in an intuitive way[46].

ONE Simulator is a widely used DTN routing simulator developed by Aalto University to support complex delay tolerant network (DTN) research and simulation. The simulator is implemented in the Java programming language, a commonly used, user-friendly and extensible programming language. As an agent-based discrete event simulator, the ONE Simulator is designed to be extremely flexible and can be used for countless different purposes. It addresses complex DTN simulation needs by integrating mobility modeling, routing simulation, visualization, and reporting. Users can choose to use the integrated mobility model on demand or import mobility data from external sources. These data are then used to determine whether the nodes can communicate and exchange information and can be used by external or internal routing modules. ONE Simulator has an interactive Graphical User Interface (GUI) that enables users to observe, and report simulated data in real time. By employing a time-slicing approach, the simulator advances the simulation at fixed time steps, allowing for efficient movement and routing simulations. Inside the simulator, wireless nodes are grouped, and each group of nodes shares a common set of parameters. This structure allows the simulator to easily simulate different scenarios, such as the interaction of pedestrians, vehicles, and public transportation. Most importantly, the ONE emulator is extremely scalable. It can dynamically load various mobile models, reporting modules, routing algorithms and event generators. This feature makes the extension and custom configuration of the simulator very simple, providing researchers and developers with a powerful and flexible tool to deeply explore and understand the multifaceted content of DTN[46]. Figure 1 summarizes how the elements in the ONE simulator interact with each other.

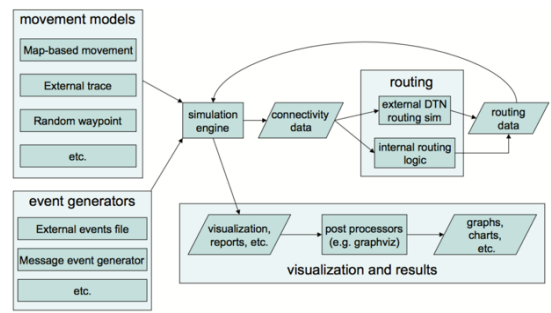

Figure 1: ONE Simulator working principal diagram.

#### 3.4.2 Simulate the implementation process

3.4.2.1 Data preprocessing and import

As mentioned earlier, the ONE simulator can use the experimental motion model of the external motion data (ExternalMovement) to read the time-stamped node positions from the file and move the nodes in the mobile simulation accordingly. By referring to the javadocs of the ExternalMovementReader class of the ONE simulator, as shown in Figure 2, the format of the externally imported data of the ONE simulator is obtained. The first line of the data format file should be the offset header, and its syntax should be: minTime maxTime minX maxX minY maxY. The rest of the data file should include a series of time and position tuples, the syntax of which should be time id xPos yPos, followed by timestamp, taxi id, longitude coordinates, and latitude coordinates.

x

```
public List<Tuple<String, Coord>> readNextMovements() {
    ArrayList<Tuple<String, Coord>> moves =
        new ArrayList<Tuple<String, Coord>>();

    if (!scanner.hasNextLine()) {
        return moves;
    }

    Scanner lineScan = new Scanner(lastLine);
    double time = lineScan.nextDouble();
    String id = lineScan.next();
    double x = lineScan.nextDouble();
    double y = lineScan.nextDouble();

    if (normalize) {
        time -= minTime;
        x -= minX;
        y -= minY;
    }

    lastTimeStamp = time;

    while (scanner.hasNextLine() && lastTimeStamp == time) {
        lastLine = scanner.nextLine();

        if (lastLine.trim().length() == 0 ||
            lastLine.startsWith(COMMENT_PREFIX)) {
            continue; /* skip empty and comment lines */
        }

        // add previous line's tuple
        moves.add(new Tuple<String, Coord>(id, new Coord(x,y)));

        lineScan = new Scanner(lastLine);
```

Figure 2: ONE Simulator External Import Data Format.

The following four steps were taken to convert the San Francisco Taxi dataset into a format acceptable to ONE SimulatorExternalMovement. Firstly, The taxi IDs are converted to integer form by using the Pandas astype('category') and 'cat.codes' methods. The goal is to efficiently encode taxi IDs as a contiguous sequence of integers. Secondly, Convert the timestamp from Unix time to a specific date by using the 'to_datetime' function of Pandas and the 'datetime' library, and limit the dataset to this day by filtering the data with the date of May 21, 2008, using the sort_value function sorts of data in ascending time order. Thirdly, By the proportional mapping method, i.e., calculate the difference of each coordinate from the minimum coordinate, divide it by the latitude and longitude range, and multiply by the width or height of the simulator. Linearly converts geographic coordinates to the simulator's coordinate system, but this conversion ignores the curvature of the Earth and therefore only works for small geographic regions. Finally, Read the modified data file, calculate the minimum and maximum timestamps, X coordinates and Y coordinates, parse the offset and range, and use Python's random.randint function to generate a simulation value for each node at the beginning of the simulation. unique random starting position in San Francisco. Finally, the processed dataset is invoked through ExternalMovement.

3.4.2.2 Introducing the urban scenario of San Francisco into the ONE simulator

By configuring 'GUI.UnderlayImage.offset' to -30, -40 to adjust the image's pixel offset in the X and Y directions, using 'GUI.UnderlayImage.scale' at 3.23 to scale the image, and setting 'GUI.UnderlayImage.rotate' to 0 to maintain the image's rotation angle, these parameters successfully integrated the San Francisco photo into the ONE simulator, achieving the visualization effect as shown in Figure 3.

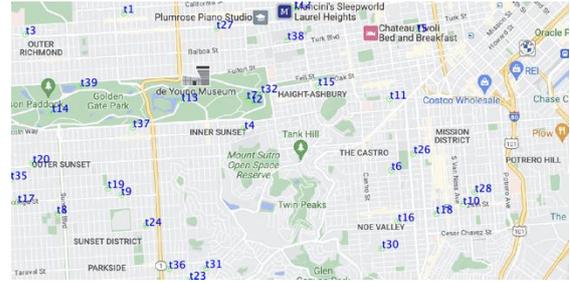

Figure 3: ONE simulator user interface diagram

3.4.2.3 Confirmation of simulation parameters of ONE simulator

In an urban environment, due to the existence of buildings and other obstacles, the communication range is limited, and the low transmission rate of Bluetooth technology is adapted to the short-distance and low-energy vehicle communication needs. At the same time, the different data transmission needs of taxis also need to be taken into account. Based on these factors, the confirmed parameters and their value ranges are as follows:

btInterface.transmitRange: The range of the interface, which is set to 10 to 30 meters, reflects the actual situation of short-distance vehicle communication in the city as much as possible, taking into account the possible impact of obstacles.

btInterface.transmitSpeed: The transmission speed of the interface, select the range from 125 to 375 KBps, suitable for data exchange between on-board devices.

Group.bufferSize: The size of the node message buffer. The buffer size of the on-board device is usually not large, so it is set to a range of 500 to 10240 MB.

Group.waitTime: Waiting time interval, the time for taxis to wait for customers is variable, and the setting range is from 60 to 900 seconds.

Group.msgTtl: The value of this parameter reflects the message delay and time-to-live characteristics in the urban communication network, and the value is 1800 to 7200 seconds.

Events.interval and Events_size: These parameters are the message propagation interval and size respectively, which can simulate continuous data streams in taxis (such as GPS updates) and occasional large data transmissions (such as video uploads), and the setting range is 5 to 15 seconds and 20 to 50k respectively.

Since the ONE simulator references an external dataset of San Francisco taxis, the variable values of nrofHosts and MovementModel are fixed. Then, run the simulation by executing ./one.sh, select Delivery_prob and Overhead_ratio as the target indicators, and perform hundreds of simulations to collect sufficient data sets, which will provide the basis for subsequent machine learning analysis.

3.4.2.4 Selection of evaluation indicators

The delivery probability is the ratio between the number of successfully delivered messages and the total number of messages sent. In the context of the taxi network in San Francisco, increasing the delivery success probability ensures that information is transmitted more reliably in an unstable environment. The delivery probability is measured using the following equation :
$$DeliveryProbability = \frac{number\ of\ messages\ received}{number\ of\ messages\ sent} \quad (1)$$

Average Latency is the average time between when a message is created and when the message was received at its destination (the duration it takes for a message to be delivered). In dynamic environments like the taxi network in San Francisco, reducing the overhead ratio can prevent energy wastage on ineffective and redundant message transmissions. The Average Latency can be measured using:

$$Average\ Latency = \sum_{i=1}^{n}\left(\frac{time\ message\ received - time\ message\ created}{number\ of\ messages\ received}\right) \quad (2)$$

In an urban environment, especially a busy metropolis like San Francisco, the reliability and efficiency of messaging is critical to many critical services. Services such as traffic, first aid, and municipal administration may rely on DTN to deliver critical information. Therefore, it is chosen to increase Delivery_prob while reducing overhead_ratio as an indicator for optimizing DTN energy consumption.

### 3.5 Introduction to Machine Learning and Modeling Optimization Process

This subsection will firstly define Random Forests and GBMs and explain how they work. We next explain why these models are suitable for this study, and finally detail how they were modeled.

#### 3.5.1 Random Forest Model

Random forest is an ensemble learning method proposed by Leo Breiman in 2000. Its formulation was influenced by Amit and Geman's (1997) on geometric feature selection, Ho's (1998) random subspace method, and Dietterich's (2000) random partition selection method. A random forest consists of multiple decision trees, where each tree depends on an independent, identically distributed random vector and is trained on a randomly chosen subspace of the data. Its core idea is that when the training data is fed into the model, the random forest builds multiple small decision trees by using different subsets and feature attributes, and then merges them into a more powerful model, rather than using the entire training Data to build a large decision tree. Random forests can enhance the performance of a model by combining the results of multiple decision trees. At the same time, each subset of the random forest is established through the samples selected by the random forest and the feature attributes selected by the random forest. This randomization reduces the sensitivity of the decision tree to the training data, thus preventing overfitting[47]. The flow chart of Random Forest is shown in the Figure4:

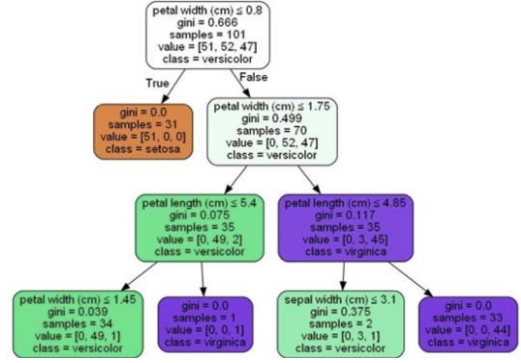

Figure 4: flow chart of Random Forest

The working principle of Random Forest can be divided into the following steps: Firstly, randomly select a subset from the training data set, and at the same time randomly select some of its feature attributes. Then, build a Decision Tree model, and use this subset and feature attributes to train the model. These two steps are repeated until the specified number of decision trees are built. When unknown data is input, each Decision Tree will make a prediction, and then according to the prediction results of all Decision Trees, the final prediction result is obtained by voting or averaging[47].

For classification problems, the prediction of Random Forest is the voting result of all trees, and each tree classifies and votes for samples, the following formula:

$$y^\wedge = argmax_i \sum j = 1_m I(f_j(x) == i) \quad (3)$$

Among them, y^ is the final classification result, fj(x) is the classification result of the jth Decision Tree, m is the number of Decision Trees, i is the indicator function, when the condition is true, return 1, otherwise return 0.

For regression problems, the average of the predictions of all trees is used. The following formula:

$$y^\wedge = m1_{j=1} \sum mf_j(x) \quad (4)$$

Among them, y^ is the final regression result, fj(x) is the regression result of the jth Decision Tree, and m is the number of Decision Trees[47].

The Random Forest is chosen as the model of DTN energy consumption for the following reasons: Firstly, the robustness of the Random Forest makes it perform well even when dealing with noisy data or outliers and is suitable for complex DTN environments. Secondly, by integrating multiple Decision Trees, the model can better balance the trade-off between the ability to fit data and maintain generalization performance, thereby reducing the risk of overfitting. Then, the Random Forest model is highly interpretable, and each Decision Node corresponds to a simple test of the input features, which can help understand which features are most influential in the case of optimizing energy consumption[47]. The most important thing is that Random Forest does not depend on the scale of features, and it can better handle the complex nonlinear relationship between parameters in the DTN environment.

### 3.5.2 GBM Model

Gradient Boosting Machine (GBM) is an ensemble learning method proposed by Jerome Friedman in 2001, following the statistical interpretation of gradient boosting algorithms by Friedman et al. in 2000. The construction idea of GBM was influenced by the boosting series algorithms, especially the gradient descent algorithm framework proposed by Freund and Schapire in 1997. GBM builds a powerful predictive model by serially training multiple weak learners (usually decision trees) and combining them[5]. The working process of GBM involves iteratively optimizing the predictive ability of the model. In each iteration, GBM firstly uses the current model to predict the data and calculates the residual between the predicted value and the actual value. This residual fitting process reflects the continuous refinement of the model. A new weak learner is then trained to fit this residual and added to the current model. By repeating this process, GBM gradually reduces the residual, thereby incrementally improving the predictive ability of the model. This methodology allows GBM to combine the strengths of multiple weak learners into a robust ensemble model that can make more accurate predictions[5].

The algorithm of gradient boosting machine can be described by the following formula:

Initialize the model: Set a constant prediction as the initial model, usually choosing the mean of the training data:

$$F_0(x) = \frac{1}{n}\sum_{i=1}^{n} y_i \quad (5)$$

Initialize the model: Set a constant prediction as the initial model, usually choosing the mean of the training data.

Iteratively train weak learners: For each iteration m=1,2,…,M, perform the following steps:

a. Calculate the negative gradient (residual):

$$r_{im} = -[\frac{\partial L(y_i, F(x_i))}{\partial F(x_i)}]_{F(x) = F_{m-1}(x)} \quad (6)$$

where L is the loss function

b. Fit residuals: Fit residuals using a weak learner (e.g., decision tree) $r_{im}$, get: $h_m(x)$

c. Calculation step size: find a value that minimizes the following losses $a_m$:

$$a_m = arg\, min \sum_{i=1}^{n} L(y_i, F_{m-1}(x_i) + ah_m(x_i)) \quad (7)$$

d. Update the model:

$$F_m(x) = F_{m-1}(x) + a_m h_m(x) \quad (8)$$

e. Get the final model:

$$F(x) = F_M(x) = F_0(x) + \sum_{m=1}^{M} a_m h_m(x) \quad (9)$$

Where $F_m(x)$ represents the model after the m-th iteration, $h_m(X)$ is the weak learner fitted in the m-th iteration, and $a_m$ is the step size of the iteration. The loss function $L(y, F(X))$ is used to measure the error between the predicted value F(x) and the actual value y[5].

The GBM is chosen as the model of DTN energy consumption for the following reasons:

Firstly, GBM is able to be tailored to the specific needs of an application, allowing the use of different loss functions and being able to handle complex nonlinear functional dependencies[5]. Secondly, the feature importance ranking provided by the GBM model helps to deeply understand the impact of different features on the prediction results. Finally, since the decision tree of GBM splits based on the ordering information of features rather than their actual values, the model is robust in terms of feature scale. This feature means that there is no need to perform feature scaling on input variables when predicting specific values, which is in line with the needs of this project to predict specific values.

### 3.6. The process of building a Random Forest Model

#### 3.6.1 Data preprocessing stage

Firstly, the dataset was cleaned, the unnecessary 'sample' column was removed, and values with specific units were converted to floats. Specific routers EpidemicRouter and MaxPropRouter in the "Group.router" column are converted to 0 and 1 respectively, while comma-separated strings in the "Group.waitTime" column are converted to integers. Next, by plotting a heatmap of the correlation matrix, we gained insight into the structure of the data and the interrelationships between features. The relationship between each feature and the target (i.e., 'delivery_prob' and 'overhead_ratio') was further analyzed by means of scatterplots, histograms and boxplots. Finally, for model training and evaluation, the dataset is divided into 80% training set and 20% test set.

#### 3.6.2 Model construction and optimazation

Firstly, a basic model of a random forest regressor was created through RandomForestRegressor, and the default parameter settings were used to fit it on the training set. Predictions are made on the test set, and evaluation metrics such as MSE, RMSE, and R-squared are calculated. Then through grid search and 5-fold cross-validation, different parameter combinations were explored, including n_estimators, max_features, max_depth, min_samples_split, min_samples_leaf and bootstrap. The optimal combination of parameters was found, and a new random forest model was created for training and evaluation. Finally, using a Feature selection method to identify features with greater influence and a new random forest model was trained based on the selected features and optimal parameter combinations. Predictions were made on the test set and model performance was evaluated.

#### 3.6.3 Forecasting the optimal parameter combination for DTN energy optimazation

Firstly, define a function named objective, which accepts a vector x containing multiple parameter values, and then encapsulates these parameter parameters into a DataFrame, and predicts through the Random Forest model to obtain delivery_prob and overhead_ratio. Then define the value range of the input variable btInterface.transmitSpeed, btInterface.transmit,RangeGroup.bufferSize,Group.waitT

ime,Group.router,Group.msgTtl,Events1.interval,Events1.size. Finally, use the differential evolution algorithm and the differential_evolution function in the scipy.optimize library to the objective function and the boundary range are used as input, and the parameter combination that minimizes the objective function is found within the given range.

### 3.7. The process of building a GBM model

#### 3.7.1 Data preprocessing stage

Firstly, clean the dataset, remove the column named "sample", and convert the values with specific units to floats. Route types in the "Group.router" column are converted to 0 and 1, while comma-separated strings in the "Group.waitTime" column are converted to integers. Next, inspect the dataset for missing values and provide a summary of descriptive statistics. Gain deeper insights into the structure and relationships between features by creating a heatmap of the correlation matrix. Further analyze the relationship between each feature and the target variable using scatter plots and violin plots. Lastly, for model training and evaluation, split the dataset into a 70% training set and a 30% test set.

#### 3.7.2 Model construction and optimization

Firstly, a GBM model is created through the GradientBoostingRegressor class in sklearn.ensemble. The model is paired by setting initial hyperparameters such as n_estimators, learning_rate and max_depth, then the model is trained on the training set, predictions are made on the training set and test set, and evaluation indicators such as MSE, RMSE and R square are calculated. Next, introduce some other hyperparameters such as subsample, min_samples_split, min_samples_leaf and max_features. Train the model again and evaluate performance using the same evaluation metric to see if these new features improve the model. Finally, hyperparameter tuning was performed using GridSearchCV. A grid of hyperparameters was defined and a grid search was performed on the entire training set using cross-validation. After finding the best combination of hyperparameters, predictions were made on the test and training sets using the best model, and the same performance metrics were calculated.

#### 3.7.3 Forecasting the optimal parameter combination for DTN energy optimazation

Firstly, define the objective function by extracting the real value and predicted value of the features "delivery_prob" and "overhead_ratio". These values are then scaled using MinMaxScaler. Subsequently, calculate the mean squared error and 1minus mean squared error to represent scores for "overhead_ratio" and "delivery_prob" respectively, then use the weights to combine the two scores into one weighted score and return it. Finally, the optimal solution is found within the given parameter range by using the differential_evolution global optimization algorithm. During optimization, predictions are made using the best model previously found by grid search, and the predictions are passed to the objective function along with a predefined target value to calculate the score.

## IV. RESULTS

### 4.1. Performance comparison between Epidemic and MaxProp

Figure 5 shows the average performance of delivery_prob using Epidemic and MaxProp protocols in the San Francisco taxi network. Among them, the delivery probability of the Epidemic protocol is 0.2135, while the MaxProp is 0.2257.

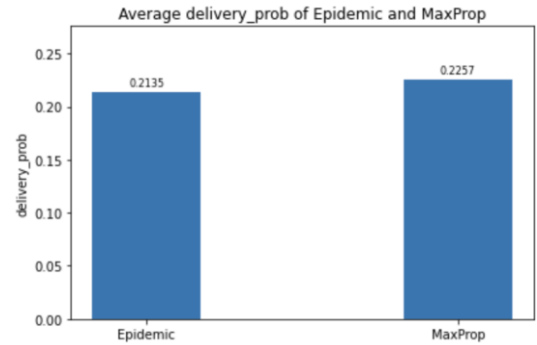

Figure 5: Average delivery_prob of Epidemic and MaxProp

Figure 6 reflects the average overhead_ratio of Epidemic and MaxProp protocols in the San Francisco taxi network. Epidemic protocol is 7.1408, while MaxProp is 5.6816.

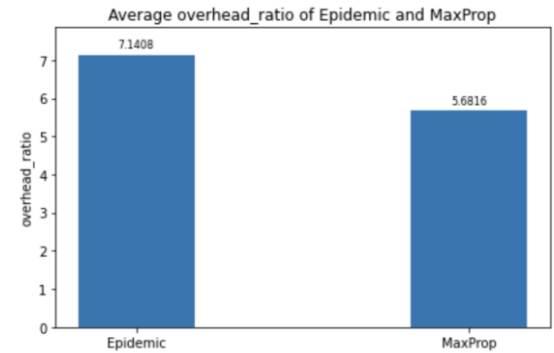

Figure 6: Average overhead_ratio of Epidemic and MaxProp

Boxplot Plot 7 shows the distribution of delivery_prob in the San Francisco taxi network using the Epidemic and MaxProp protocols. For the Epidemic protocol, the median delivery_prob is 0.1925, representing the value below which half of the delivery_prob observations fall, while the other half are above it. Q1 is 0.07375, indicating that 25% of the delivery_prob values are below this point. Q3 is 0.32775, indicating that 75% of the delivery_prob values are below this point. The minimum delivery_prob is 0.0067, and the maximum is 0.6493, showcasing the lowest and highest delivery_prob values achieved by the Epidemic protocol across all simulations.

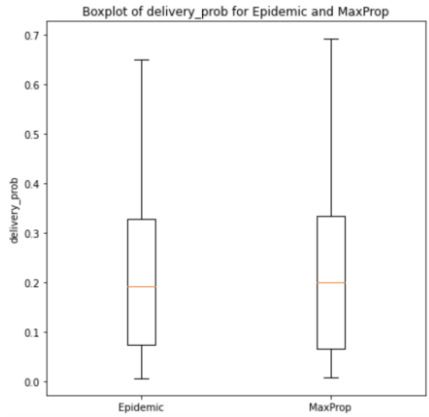

Figure 7: Boxplot of delivery_prob for Epidemic and MaxProp

Regarding the MaxProp protocol, the median delivery_prob is 0.2005, with Q1 at 0.0662 and Q3 at 0.3346. The minimum delivery_prob is 0.0085, and the maximum is 0.6924. As shown in Table 1.

Table 1: Boxplot of delivery_prob for Epidemic and MaxProp

|         | Epidemic | MaxProp |
|---------|----------|---------|
| median  | 0.1925   | 0.2005  |
| Q1      | 0.07375  | 0.0662  |
| Q3      | 0.32775  | 0.3346  |
| minimum | 0.0067   | 0.0085  |
| maximum | 0.6493   | 0.6924  |

Boxplot Plot 8 shows the distribution of overhead_ratio in the San Francisco taxi network using the Epidemic and MaxProp protocols. For the MaxProp protocol, the median overhead_ratio is 3.8645, signifying that half of the observations have an overhead_ratio value lower than this, while the other half have a higher value. Q1 is 3.4461, indicating that 25% of the overhead_ratio values fall below this point. Q3 is 6.7188, indicating that 75% of the overhead_ratio values fall below this point. The minimum overhead_ratio is 2.0509, and the maximum is 21.377, illustrating the lowest and highest overhead_ratio values achieved by the MaxProp protocol across all simulations.

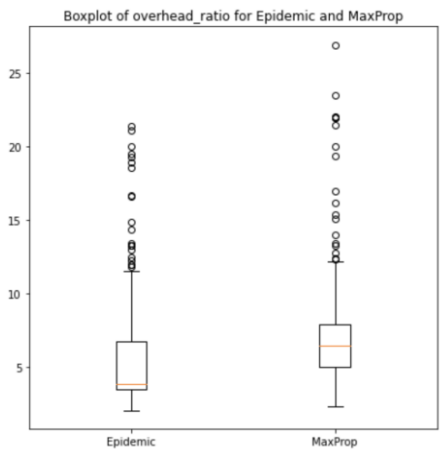

Figure 8: Boxplot of overhead_ratio for Epidemic and MaxProp

Concerning the Epidemic protocol, the median overhead_ratio is 6.4824, with Q1 at 4.98385 and Q3 at 7.90475. The minimum overhead_ratio is 2.2938, and the maximum is 26.9444. As shown in Table 2:

Table 2: Boxplot of overhead_ratio for Epidemic and MaxProp

|         | MaxProp | Epidemic |
|---------|---------|----------|
| median  | 3.8645  | 6.4824   |
| Q1      | 3.4461  | 4.98385  |
| Q3      | 6.7188  | 7.90475  |
| minimum | 2.0509  | 2.2938   |
| maximum | 21.377  | 26.9444  |

The line graph 9 shows that the delivery_prob of the MaxProp protocol is generally higher than that of the Epidemic protocol under different simulation times. This indicates that packets using the MaxProp protocol are more likely to be successfully delivered, which may lead to lower energy consumption.

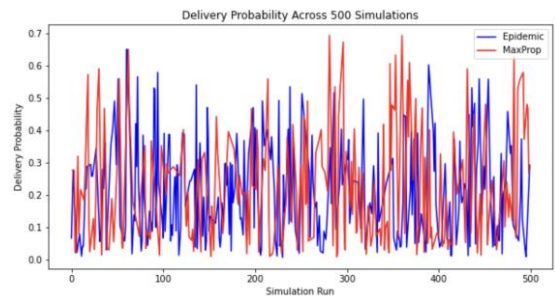

Figure 9: Trends in delivery_prob: 500 Runs of Epidemic vs. MaxProp

The line chart 10 indicates that, when comparing overhead_ratio, the overall trend of the Epidemic protocol is higher than that of the MaxProp protocol. A higher overhead_ratio suggests that the Epidemic protocol generates more network overhead when delivering packets, resulting in increased network energy consumption. Comparing Figure 9, it can be found that In the MaxProp protocol, higher delivery_prob points are often accompanied by higher overhead_ratio.

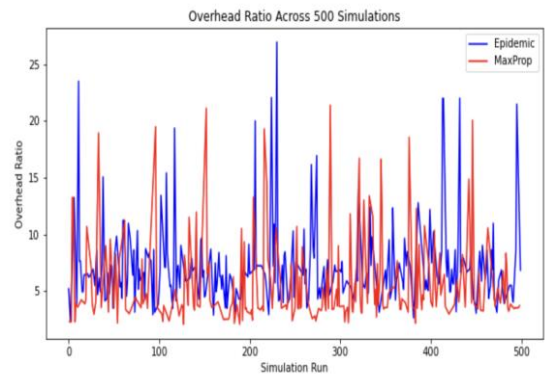

Figure 10: 500 Simulations Overview: overhead_ratio Patterns for Epidemic and MaxProp

From the above histograms, boxplots, and line graphs, it can be seen that the MaxProp protocol is superior to the Epidemic protocol in terms of performance. This observation comes as no surprise. AS Burgess et al[3], introduced the MaxProp protocol as a response to the energy-constrained challenges of vehicle-based Delay Tolerant Networks. This protocol focuses on achieving efficient routing optimization in complex environments. By carefully designing the replication mechanism of data packets, MaxProp not only successfully improves the probability of delivery, but also optimizes energy consumption and avoids unnecessary excessive replication. Compared with this, the performance of Epidemic protocol in these aspects is relatively weak. Therefore, if you want to find a balance between overhead_ratio and deliver_prob, the MaxProp protocol may be a more suitable choice.

### 4.2. Sensitivity analysis of multiple parameters

In this section, a series of visual graphs will be used to show the relationship between the various parameters involved in the research.

**4.2.1 Heat Map**

Figure 11 provides a detailed presentation of the correlation between various input parameters and the target variable (i.e., delivery_prob and overhead_ratio) in the ONE simulator. The analysis reveals that the influence on delivery_prob is ranked in descending order as follows: btInterface.transmitRange (0.68), btInterface.transmitSpeed, Events1.size, Events1.interval, Group.msgTtl, Group.bufferSize, Group.router, and Group.waitTime (-0.035). As for the impact on overhead_ratio, the most significant parameter is btInterface.transmitSpeed (-0.4), followed by Events1.size, btInterface.transmitRange, Group.router, Events1.interval, Group.waitTime, Group.msgTtl, and Group.bufferSize (0.046).

The relationship between btInterface.transmitRange and btInterface.transmitSpeed is weak, with a coefficient of 0.035. This might be since in urban environments, the correlation between signal transmission distance and speed is not always proportional. For instance, tall buildings might obstruct signals, leading to a less obvious connection between transmission speed and distance. Additionally, Interface.transmitRange exhibits very low correlations with Group.bufferSize and Group.waitTime, measuring -0.0018 and 0.019, respectively. This suggests that in urban scenarios, an increase in transmission range doesn't necessarily lead to greater buffering requirements or longer waiting times, as other factors like traffic congestion or signal interference may also play a role.

Further analysis revealed a notable correlation (0.28) between Events1.size and overhead_ratio. This could be attributed to significant events in urban settings, such as traffic accidents or public gatherings, placing additional strain on communication and transportation systems. For instance, road congestion resulting from unforeseen events might require traffic rerouting, thereby increasing system overhead.

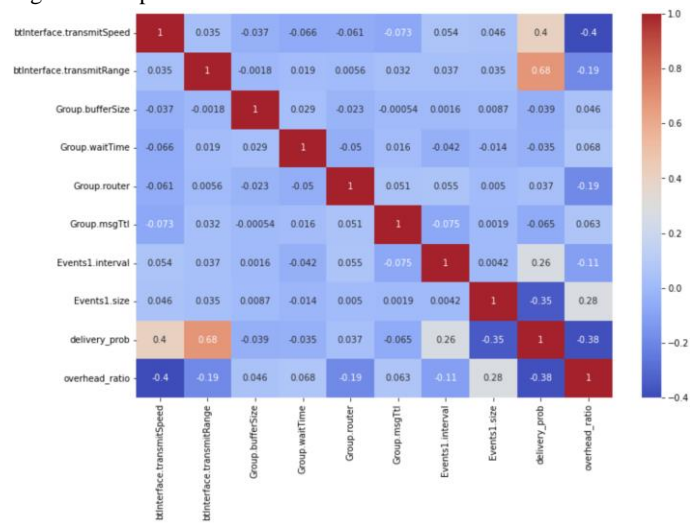

Figure 11: Correlation Heatmap of Input Variables in the ONE Simulator

**4.2.2 Scatter Plots**

Figure 12 presents a scatter plot revealing the relationships between various parameters and delivery_prob. Firstly, it's evident from the graph that there is a significant positive correlation between btInterface.transmitSpeed and btInterface.transmitRange with delivery_prob. As these two parameters increase, delivery_prob also rises. This means that increasing transmission speeds and extending transmission range may help increase the probability of successful transmissions, possibly due to stronger signals and wider coverage.

On the contrary, Group.buffersize does not exhibit a clear correlation with delivery_prob. This implies that in certain scenarios, increasing buffer size might not directly lead to an improvement in transmission success rates.

At the same time, the data points for Group.waittime and Events1.interval appear to be clustered, which may indicate that certain events, such as traffic accidents or traffic signal changes, occur at consistent time intervals. Especially during rush hour, this could result in a noticeable increase in waiting times.

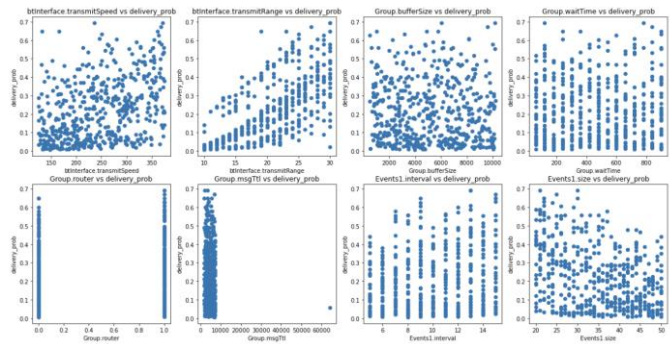

Figure 12: Cross-Layer Correlation Scatter Plot: Parameter Interdependencies Influencing Delivery_prob

Figure 13 provides an in-depth exploration of the relationships between various parameters and overhead_ratio. It is observed that when the btInterface.transmitSpeed is lower than 200k, the overhead_ratio has an obvious upward trend, which may be caused by network congestion. Additionally, with the increase of btInterface.transmitRange, overhead_ratio exhibits a declining trend, indicating that expanding the transmission range can enhance communication efficiency and reduce the demand for network resources.

Another interesting finding is that there is a positive correlation between Events1.size and overhead_ratio, which may be related to the pressure of large-scale events on the network. At the same time, the distribution of Group.waittime and Events1.interval in overhead_ratio is similar to delivery_prob.

It is worth noting that despite the change in Group.buffersize, overhead_ratio still increases, which means that the impact of Group.buffersize on overhead_ratio may not be obvious.

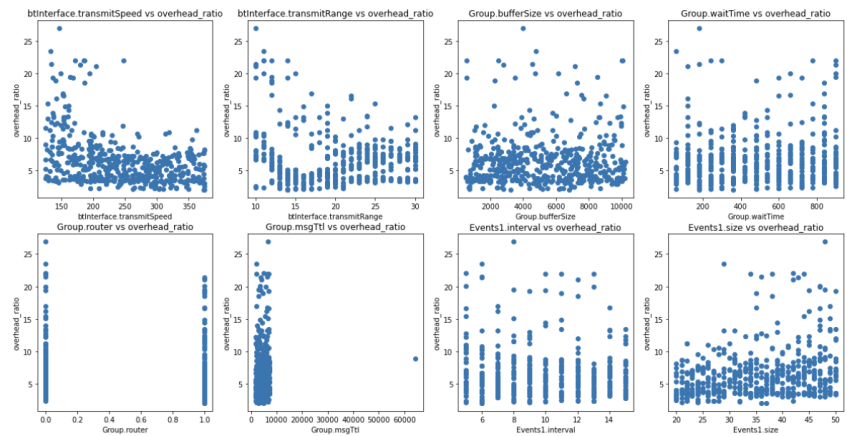

Figure 13: Interdependency Analysis: Scatter Plot of Parameters vs Overhead_Ratio

### 4.2.3 Violin Plots

Figure 14 illustrates that when btInterface.transmitRange is in the range of 28 to 30 meters, the impact on delivery_prob is most obvious. This might reveal a common phenomenon in communication networks: as the transmission range expands, the number of accessible nodes increases, enhancing the potential for information transmission and consequently elevating the probability of successful transmission.

Conversely, when Group.time falls within the interval of 480 to 600 seconds, delivery_prob exhibits a decreasing trend. This could suggest that within this specific time interval, factors like network congestion or other influences might affect the success rate of transmission.

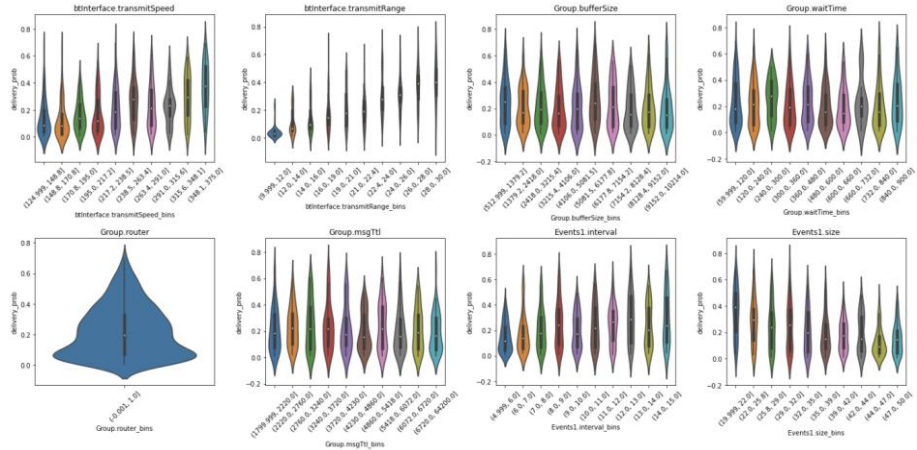

Figure 14: Multi-layer Violin Plot: Cross-layer Correlation of Parameters with delivery_prob

Figure 15 illustrates the correlation between different parameters and overhead_ratio through the volin diagram. The Figure clearly shows that the overhead_ratio reaches its peak when the btInterface.transmitSpeed is 125 to 149k and the btInterface.transmitRange is 10-12 meters. It may be that this speed is not able to effectively cover the complex environment composed of tall buildings, traffic and crowds in the city.

The btInterface.transmitRange of 10-12 meters may be insufficient, resulting in the signal not being able to penetrate obstacles or bypass interference sources.

Therefore, more resources may be required to ensure the success of the communication, thus increasing the overhead_ratio. When the Events1.size is 20 to 22k, the overhead_ratio is the smallest.

For typical mobile communication, IoT devices and smart city services in the city, this size may neither cause fragmentation and slow transfer of data, nor be too small to increase the burden of frequent communication and processing. This can be an ideal packet size in application scenarios such as intelligent traffic signal control, environmental monitoring, etc.

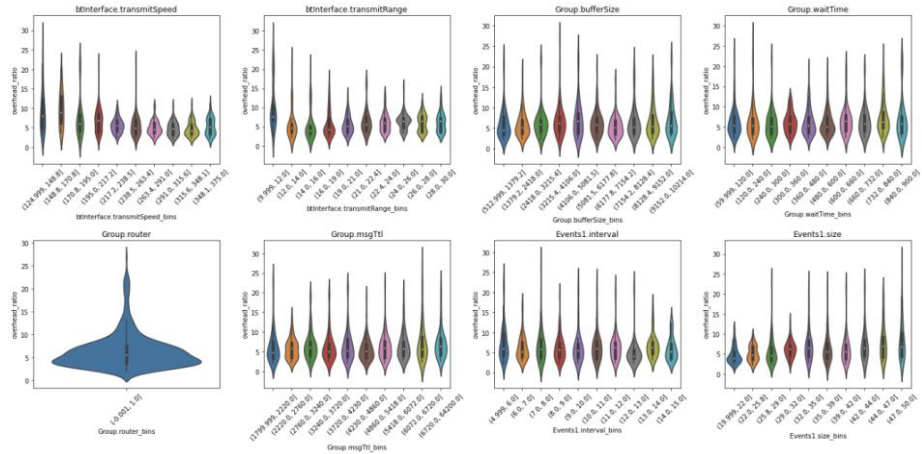

Figure 15: Violin Visualization of Parameter Interdependencies and Their Impact on overhead_ratio

### 4.2.4 Box plots

Figure 16 presents a boxplot illustrating the interrelationships between different parameters and delivery_prob. This visualization provides insights into the data distribution of each parameter and possible outliers, contributing to a deeper understanding of how these parameters influence delivery_prob.

Firstly, it is worth noting that when the transmission range is located between 28 and 30 meters, we observe a clear outlier of about 0.05. Although delivery_prob values in this range are generally high, this outlier may suggest that in some specific cases, external factors such as signal attenuation or interference may significantly reduce delivery_prob.

Furthermore, a significant outlier of approximately 0.68 appears when the buffer size falls within the range of 9244 to 10214 MB. This anomaly is not solely determined by buffer size alone; other parameters, such as transmission speed and range, could also play a role.

These observations further highlight the need for comprehensive tuning and optimization of multiple parameters in order to effectively improve delivery_prob and simultaneously reduce overhead_ratio in a DTN environment.

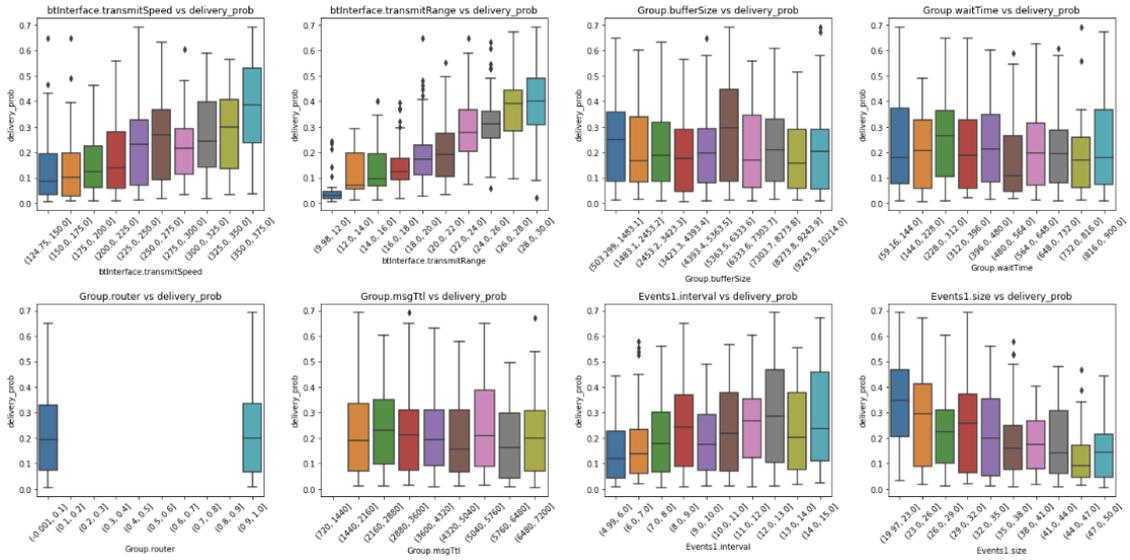

Figure 16: Cross-layer Correlation Boxplot: Parameters Impacting Delivery Probability

Figure 17 depicts the relationships between different parameters and overhead_ratio through a boxplot. Clearly visible from the graph, there are some notable outliers that stand out above the rest when Events1.interval is between 7 and 8s. These prominent data points may represent instances where data request volumes sharply increased due to specific events within these time intervals, such as large-scale public activities or urgent situations, thereby causing a surge in communication system overhead.

Additionally, when Events1.size falls within the range of 47 to 50k, some higher outliers are also present. These values suggest a strong demand for large data transmission within this data size range, potentially related to scenarios where substantial data transmission is required, such as high-definition video streaming or large-scale real-time live events.

These findings highlight the importance of adapting and optimizing resource allocation and strategies to suit different communication needs and scenarios.

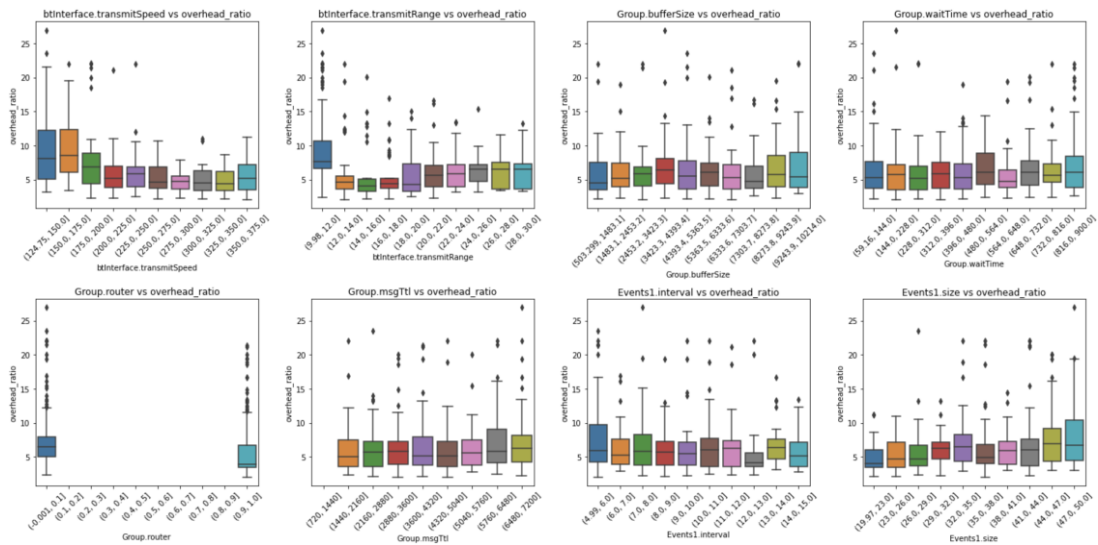

Figure 17: Cross-layer Interdependency Analysis: Boxplot Visualization of Parameters Influencing Overhead Ratio

### 4.3. Performance of Machine Learning Models

For the Machine Learning part, two models were chosen: Random Forest and Gradient Boosting Machine (GBM). Both models are used for regression prediction.

#### 4.3.1 Default Parameter

Random forest run with default parameters, n_estimators=50, max_features='log2, max_depth=12, min_samples_split=20, min_samples_leaf=20, bootstrap=True. The results are as shown in Table 3:

Table 3: Performance of Random Forest Model with Default Parameters

|  | Train set | Test set |
|---|---|---|
| Mean Squared Error | 3.53 | 5.73 |
| Root Mean Squared Error | 1.88 | 2.39 |
| R-squared | 0.51 | 0.42 |

The GBM model adopts default parameters, n_estimators=100, learning_rate=0.04, max_depth=2, subsample=0.1,min_samples_split=10,min_samples_leaf =6, max_features='auto'. The running results are as shown in Table 4:

Table 4: Performance of GBM Model with Default Parameters

|  | Train set | Test set |
|---|---|---|
| Mean Squared Error | 3.41 | 4.1 |
| Root Mean Squared Error | 1.84 | 2.02 |
| R-squared | 0.67 | 0.63 |

From the above results, it can be observed that the Gradient Boosting Machine performs slightly better than the Random Forest on the test set, especially on the R-squared value. This means that Gradient Boosting Machine models may be better suited for this dataset and prediction task.

#### 4.3.2 Grid Search

Through Grid Search, The Random Forest Model obtained the optimal parameters: bootstrap: True, max_depth: 10, max_features: log2, min_samples_leaf: 15, min_samples_split: 10, n_estimators: 15. Upon retraining the model with these parameters, the results are as shown in Table 5:

Table 5: Performance of Random Forest Model with Grid Search

|  | Train set | Test set |
|---|---|---|
| Mean Squared Error | 3.22 | 5.46 |
| Root Mean Squared Error | 1.79 | 2.33 |
| R-squared | 0.56 | 0.48 |

Through Grid search, The GBM Model identified the optimal parameters as follows: learning_rate: 0.04, max_depth: 2, max_features: 'auto', min_samples_leaf: 5, min_samples_split: 8, n_estimators: 100, and subsample: 0.12. After retraining the model with these parameters, the results are as shown in Table 6:

Table 6: Performance of GBM Model with Grid Search

|  | Train set | Test set |
|---|---|---|
| Mean Squared Error | 2.70 | 3.89 |
| Root Mean Squared Error | 1.65 | 1.97 |
| R-squared | 0.72 | 0.65 |

After Grid Search optimization, it can be observed that after using the optimized parameters, the performance of both models improves. Especially on the R-squared value, the Gradient Boosting Machine model performs better.

#### 4.3.3 Feature Selection

In the Random Forest Model, based on the feature selection strategy, the following five features are selected: btInterface.transmitSpeed, btInterface.transmitRange, Group.router, Events.interval, and Events.size. After retraining the model with these selected features, the results are as showin in Table 7:

Table 7: Performance of Random Forest Model with Feature Selection

|  | Train set | Test set |
|---|---|---|
| Mean Squared Error | 2.85 | 5.13 |
| Root Mean Squared Error | 1.69 | 2.27 |
| R-squared | 0.62 | 0.53 |

In the GBM model, through the feature selection strategy, the following five features are selected: Events.size, btInterface.transmitSpeed,Group.router,btInterface.trans mitRange, and Group.bufferSize. After retraining the model with these selected features, the results are as shown in Table 8:

Table 8: Performance of GBM Model with Feature Selection

|  | Train set | Test set |
|---|---|---|
| Mean Squared Error | 2.52 | 3.83 |
| Root Mean Squared Error | 1.59 | 2.0 |
| R-squared | 0.74 | 0.67 |

After Feature Selection, the performance of both models on Mean Squared Error, Root Mean Squared Error and R-squared has improved. This suggests that focusing on those parameters with greater impact may help further optimize the Energy Consumption of DTN.

### 4.4. Optimal parameter combination predicting

Using the optimal parameters after Grid Search to train the Random Forest model to predict the input variables: the optimal combination of btInterface.transmitSpeed, btInterface.transmitRange,Group.bufferSize,Group.waitT ime, Group.router, Group.msgTtl, Events1.interval, Events1.size is shown in Table 9:

Table 9: Optimal parameter values for Random Forest prediction

| btInterface.transmitSpeed | 306k |
|---|---|
| btInterface.transmitRange | 25m |
| Group.bufferSize | 5583M |
| Group.waitTime | 640s |
| Group.router | MaxProp |
| Group.msgTtl | 3076s |
| Events1.interval | 14s |
| Events1.size | 24k |

Put these parameters into the ONE simulator for simulation, and the results are as follows:

Table 10: Optimal parameters running results

| Delivery_prob | Overhead_ratio |
|---|---|
| 0.59 | 3.95 |

As mentioned earlier, based on the 500 simulation experiments, it is evident that higher delivery_prob leads to a higher overhead_ratio. The delivery_prob in this simulation exceeds the Q3 value of MaxProp's delivery_prob, while the overhead_ratio remains below the Q1 value of MaxProp's overhead_ratio. This effectively strikes a balance between delivery_prob and overhead_ratio, consequently achieving the goal of reducing DTN energy consumption.

The GBM model was trained with the best parameters obtained through Grid Search. The Model was then used to predict the best combination of input variables. The optimal parameter combination is shown in Table 10:

Table 11: Optimal parameter values for GBM prediction

| btInterface.transmitSpeed | 286k |
|---|---|
| btInterface.transmitRange | 29m |
| Group.bufferSize | 10188M |
| Group.waitTime | 872s |
| Group.router | MaxProp |
| Group.msgTtl | 6729s |
| Events1.interval | 14s |
| Events1.size | 24k |

Put these parameters into the ONE simulator for simulation, and the results are as follows:

Table 12: Optimal parameters running results

| Delivery_prob | Overhead_ratio |
|---|---|
| 0.66 | 4.69 |

When using the GBM model for prediction, its optimal combination of input variables can simulate a better delivery_prob value than the Random Forest model. Although the overhead_ratio has increased when using the GBM model, this increase is still acceptable because its value is still lower than the Q1 value of the overhead_ratio of the MaxProp protocol. Similar to the Random Forest model, the GBM model aims to reduce DTN energy consumption by identifying a balance point between delivery_prob and overhead_ratio.

### 4.5. Overview

In the Performance comparison between Epidemic and MaxProp section, a comprehensive visual comparison was conducted through bar plots, box plots, and line graphs, revealing the superiority of the MaxProp protocol over the Epidemic protocol in various scenarios.

Subsequently, in the Sensitivity analysis of parameters section, a further analysis of key parameters such as btInterface.transmitSpeed, btInterface.transmitRange, and Group.buffersize was conducted to assess their impact on DTN performance. Scatter plots, violin plots, and box plots provide readers with an intuitive display of the relationships between parameters and delivery_prob and overhead_ratio.

In the Performance of Machine Learning Models section, a comparison was drawn between two major machine learning models, Random Forest and GBM, under different settings such as Default Parameters, Grid Search, and Feature Selection. The research results unequivocally demonstrate the subtle advantages of the GBM model in prediction.

Finally, in the Predict the optimal parameter combination section, the model optimized through grid search was utilized to predict the optimal parameter combination, which was subsequently validated in the ONE simulator. These simulation results confirm that the predicted parameters can achieve a reasonable balance between delivery_prob and overhead_ratio, thereby optimizing DTN energy consumption.

## V. DISCUSSION

This research focuses on optimizing the Energy Consumption of DTN, especially in scenarios where the network connection is unstable or intermittently disconnected. In order to achieve this goal, the study selected two main routing protocols: Epidemic and MaxProp, and simulated it based on the real data set of San Francisco taxi mobility traces. Through the ONE simulator, this study conducted hundreds of simulation experiments and evaluated with two key indicators: delivery_prob and overhead_ratio. The optimization of these indicators aims to reduce the overhead ratio while improving the transmission success rate, thereby reducing the overall Energy consumption of DTN.

In order to achieve this optimization goal, two Machine Learning Models, Random Forest and GBM, were further adopted in this study. These models successfully predict the optimal combination of various network parameters, including:btInterface.transmitSpeed,btInterface.transmit Range, Group.bufferSize, Group.waitTime, Group.router, Group.msgTtl, Events1.interval, and Events1.size. By re-inputting these predicted optimal parameters into the ONE simulator for verification, the experimental results show that: under the influence of these optimal parameters, the transmission success rate has improved, and the overhead ratio has also been effectively reduced, further confirming the effectiveness of this method in reducing DTN Energy Consumption.

### 5.1. Performance comparison between Epidemic and MaxProp

**5.1.1 Discussion of delivery_prob of Epidemic and MaxProp protocols**

**a**. Although the average performance of the MaxProp protocol in terms of delivery_prob (0.2257) is only slightly better than that of the Epidemic protocol (0.2135), this difference is not enough to show that MaxProp is significantly better than Epidemic in all aspects. Looking further at the Q3 value, the MaxProp protocol and the Epidemic protocol are 0.3346 and 0.32775 respectively, which also shows that the two have similar performances in terms of successful packet delivery. This similar average delivery_prob and Q3 values may imply that both protocols are somewhat effective at ensuring successful packet delivery. In particular, Jindal et al.[33] pointed out that Epidemic algorithms are potentially efficient solutions for propagating information in large-scale and dynamic systems. They were easy to deploy, robust, and highly resilient to failures.

However, upon further examination of the distribution of the maximum delivery_prob values, the results indicate that the performance of MaxProp (0.6924) significantly outperforms Epidemic (0.6493). This observation aligns with the findings of Burgess et al.[3] who pointed out that the MaxProp protocol was designed to address energy-constrained challenges in VDTNs. In comparison to other protocols, the MaxProp protocol, through carefully designed mechanisms for packet replication and dissemination, proved to enhance the success rate of packet transmission and optimize energy consumption.

**b.** In the line chart 9, it can be observed that MaxProp shows a higher delivery_prob of 0.66 when the number of simulations reaches about 480. At this time, Parameter configuration is shown in Table 13:

Table 13: The MaxProp parameter is configured when the number of simulations is 480

| btInterface.transmitSpeed | 364k |
|---|---|
| btInterface.transmitRange | 29m |
| Group.bufferSize | 513M |
| Group.waitTime | 600s |
| Group.msgTtl | 3720s |
| Events1.interval | 9s |
| Events1.size | 24k |

In contrast, when the configuration parameters are as shown in Table 14, the delivery_prob is only 0.03.

Table 14: Configuration Parameters and Corresponding Delivery_prob Values

| btInterface.transmitSpeed | 207k |
|---|---|
| btInterface.transmitRange | 10m |
| Group.bufferSize | 2318M |
| Group.waitTime | 240s |
| Group.msgTtl | 5520s |
| Events1.interval | 13s |
| Events1.size | 27k |

This significant difference may indicate that for a busy urban network like San Francisco, a higher btInterface.transmitSpeed (364k and 207k) and a wider btInterface.transmitRange (29m and 10m) may be more helpful for packets to transmit effectively spread in the network. Secondly, a smaller Group.bufferSize (513M and 2318M) may mean that the network can process and transmit packets more efficiently, thereby indirectly avoiding packet delays or losses due to excessively large buffers. Also, shorter Group.msgTtl (3720s and 5520s) and Events1.interval (9s and 13s) may also help packets travel across the network and reach their destination faster.

**c.** From the line chart 9, it can be observed that when the number of simulations is about 440, the delivery_prob of the Epidemic protocol reaches 0.56, and the parameter configuration is shown in Table 15.

Table 15: Parameter Configuration for Epidemic Protocol at 440 Simulations

| btInterface.transmitSpeed | 374k |
|---|---|
| btInterface.transmitRange | 30m |
| Group.bufferSize | 645M |
| Group.waitTime | 780s |
| Group.msgTtl | 4020s |
| Events1.interval | 8s |
| Events1.size | 35k |

Relatively, when the Parameter Configuration is as shown in Table 16, delivery_prob is only 0.021.

Table 16: Configuration Parameters

| btInterface.transmitSpeed | 153k |
|---|---|
| btInterface.transmitRange | 30m |
| Group.bufferSize | 8588M |
| Group.waitTime | 420s |
| Group.msgTtl | 2880s |
| Events1.interval | 14s |
| Events1.size | 49k |

The performance of the Epidemic protocol is similar to the MaxProp protocol just discussed, but it is worth noting that when btInterface.transmitRange is the same, btInterface.transmitSpeed and Group.bufferSize may have played a more important role. In addition, these results may also be affected by changes in the actual network and network requirements within a particular region. For example, higher transmission speeds and smaller buffers may provide higher delivery probabilities in high-demand areas of the city center, while these factors may have less impact in suburban areas with relatively less traffic. This emphasizes the importance of optimizing network parameters according to specific application scenarios.

**d.** From the parameter sensitivity analysis reflected in graphs 11, 12, and 13, a notable positive correlation emerges between btInterface.transmitSpeed, btInterface.transmitRange, and delivery_prob. This observation holds true for both the MaxProp and Epidemic protocols. However, the MaxProp protocol benefits more from this correlation, leading to its improved delivery_prob performance. However, Graph 9 dispels our earlier assumption about the linear relationship between Group.bufferSize and delivery_prob. Graph 9 indicates that the buffer size does not exhibit a linear correlation with delivery_prob. Contrary to the previous inference, a smaller buffer size does not necessarily result in higher data packet transmission efficiency or lower packet latency. This suggests that the effect of buffer size is more nuanced and might be influenced by other network parameters or conditions. Overall, MaxProp usually outperforms the Epidemic protocol in terms of delivery_prob. The most crucial factors influencing this metric are btInterface.transmitSpeed and btInterface.transmitRange.

### 5.1.2 Discussion of overhead_ratio of Epidemic and MaxProp protocols

**a.** When analyzing the overhead_ratio of the Epidemic and MaxProp protocols, it is evident that from an average perspective, the overhead_ratio of the MaxProp protocol (5.6816) is actually lower than that of the Epidemic protocol (7.1408). This suggests that MaxProp might be more energy-efficient in terms of network resource utilization. To further validate this result, a closer examination of the median and quartiles is conducted. Specifically, the median overhead_ratio of the MaxProp protocol (3.8645) is significantly lower than that of the Epidemic protocol (6.4824), and this trend is even more obvious in the Q1 value, where MaxProp is 3.4461, while Epidemic is 4.98385. This indicates that in over half of the simulations, MaxProp indeed incurs lower costs. In contrast, at Q3, both protocols exhibit similar performance (MaxProp at 6.7188, Epidemic at 7.90475), suggesting that there is no significant difference between them in scenarios with higher overhead_ratio. This phenomenon aligns with the findings of Mishu Das et al. [18], who proposed that while MaxProp can effectively enhance data packet delivery probability, it might lead to higher costs in certain situations.

**b.** Some interesting results were found when comparing two different configurations of the Epidemic protocol. Under the first configuration, btInterface.transmitSpeed: 147k, btInterface.transmitRange: 10m, Group.bufferSize: 3988M, Group.msgTtl: 6720s, Events1.size: 48k, the overhead_ratio reached the highest 26.9444. However, under the second configuration, even if the btInterface.transmitSpeed is increased to 195k and the btInterface.transmitRange is extended to 15m, the overhead_ratio is reduced to the lowest 2.2938. I originally thought it would be an increase, because the expansion of transmission speed and transmission range may increase the transmission rate and thus increase Energy Consumption. In fact, the Group.bufferSize of the second configuration is 1500M, the Group.msgTtl is 2700s, and the Events1.size is 32k, which is much lower than the first configuration. Therefore, it is preliminarily judged that Group.buffersize, Group. msgTtl and Events1.size are the most important factors affecting overhead_ratio.

In the MaxProp protocol, the observed trend further supports that Group.bufferSize, Group.msgTtl and Events1.size are the main factors affecting overhead_ratio. Under the parameter setting of btInterface.transmitSpeed is 186k, btInterface.transmitRange is 12m, Group.bufferSize is 7283M, Group.msgTtl is 2940s and Events1.size is 38K, the overhead_ratio reached a high of 18.5472. Upon adjusting the parameters, specifically by increasing the btInterface.transmitSpeed to 298k, decreasing btInterface.transmitRange to 11m, shrinking the Group.bufferSize to 661M, reducing Events1.size to 22K, and trimming Group.msgTtl to 2040M, there is a notable reduction in overhead_ratio, bringing it down to 2.395.Therefore, in the Epidemic and MaxProp protocols, Group.bufferSize , Group.msgTtl and Events1.size are the main influencing factors of overhead_ratio.

**c.** In parameter sensitivity analysis, heatmap8 and scatterplot10 provide insight into the key parameters affecting overhead_ratio. Although initial observations suggest that Events1.size, Group.bufferSize and Group.msgTtl are the main contributing factors, it is surprising that btInterface.transmitSpeed has the most significant impact. This point is further confirmed in the scatterplot 10, especially when the transmission speed is lower than 200k, the overhead_ratio increases significantly, which may lead to network congestion, packet delay and communication speed reduction.

Therefore, the strategy of optimizing overhead_ratio needs to be reconsidered. In addition to paying attention to parameters such as Events1.size, Group.bufferSize and Group.msgTtl, the impact of btInterface.transmitSpeed also needs to be considered, especially when the transmission speed is low. This is especially important to avoid network congestion and improve overall network performance.

In summary, across several performance metrics, the MaxProp protocol generally outperforms the Epidemic protocol. However, it is important to note that while MaxProp tends to enhance delivery_prob, it often leads to a corresponding increase in overhead_ratio. Therefore, in order to reduce the energy consumption of DTNs, a balance must be struck to enhance delivery_prob while minimizing overhead_ratio as much as possible.

## 5.2. Performance Comparison of Machine Learning Models

### 5.2.1 Model Performance

The MSE and R-squared values of the GBM model in each stage are better than the Random Forest model on the test set. Nickalus Redell[48].showed that R-squared measures the proportion of variance in the dependent variable explained by the independent variables in a regression model. Meanwhile, MSE measures the average squared difference between the predicted and actual values. It provides a measure of the model's predictive accuracy, with lower values indicating better performance. Therefore, lower MSE and higher R-squared values usually indicate better predictive accuracy of the model. However, this does not directly mean that the GBM model is more effective in practical applications, i.e., in terms of optimization of DTN energy consumption. The actual verification results show that although the parameter combination predicted by the GBM model can achieve a higher delivery_prob of 0.66, its overhead_ratio is 4.69 and it also increases accordingly, compared with the parameter combination predicted by the Random Forest model of delivery_prob of 0.59 and overhead_ratio of 3.95, and it cannot be definitively said which model optimizes DTN energy consumption more efficiently.

### 5.2.2 Parameter optimazation

Grid search has indeed demonstrated its effectiveness in parameter optimization, resulting in performance improvements for both models, particularly in terms of R-squared values. However, this doesn't necessarily imply the potential for further reduction in DTN energy consumption. The reason is that models are built upon historical data for prediction, and if the actual application environment significantly deviates from the training data, the model's predictions might become inaccurate. As a result, as mentioned by Fahimeh Ghasemi et al.[49]there is no guarantee that the same level of performance will be maintained when applied to new data.

### 5.2.3 Feature selection

Parameter sensitivity analysis provides important insights when exploring whether feature selection can reduce the Energy Consumption of DTN. This analysis shows that each parameter has an impact on delivery_prob and overhead_ratio. In theory, each parameter can affect the energy efficiency of the DTN network. Therefore, it may be unreasonable to only focus on important features to reduce DTN Energy Consumption. To demonstrate this, adjustments were made to the optimal parameter combination predicted using GBM. The model placed particular emphasis on five key features: Events1.size, btInterface.transmitSpeed, Group.router, btInterface.transmitRange, and Group.bufferSize. While keeping these five features constant, the three parameters of Group.msgTtl, Group.waitTime and Events1.interval were adjusted to the median assignment of the default range. As shown in Table 17:

Table 17: Original parameter combination vs adjusted parameter combination

| Parameter | original | adjusted |
| --- | --- | --- |
| btInterface.transmitSpeed, | 286k | 286k |
| btInterface.transmitRange, | 29m | 29m |
| Group.bufferSize | 10188M | 10188M |
| Group.waitTime | 872s | 450s |
| Group.msgTtl | 6729s | 3600s |
| Events1.size | 24k | 24k |
| Events1.interval | 14s | 8s |

The results indicate that while the overhead_ratio decreased from 4.69 to 3.34, possibly reducing Energy Consumption, correspondingly, the delivery_prob also decreased from 0.66 to 0.5672. Therefore, the notion that overall Energy Consumption in DTN can be reduced solely by focusing on specific "important" features is not valid.

### 5.3. Model Usability

Both the Random Forest and GBM models successfully predicted parameter combinations that could simultaneously improve delivery_prob and reduce overhead_ratio, effectively optimizing DTN Energy Consumption. However, these models have limitations in terms of applicability and generalization. Chicco et al[50]. pointed out that the R-squared value is a statistic that measures the goodness of fit of a model, representing the percentage of variability in the target variable explained by the model. The R-squared of the Random Forest model is 0.53, while the R-squared of the GBM model is 0.65. The R-squared value of the model is not high enough, which means that the model may not capture some important factors affecting the target variable, and these uncaptured factors may have higher importance in different DTN network application scenarios. If these unconsidered factors play a major role in other types or scales of DTN networks, the generalization ability of the model in these new scenarios may be greatly reduced. Furthermore, these models were trained on actual dataset from San Francisco taxi mobility traces, indicating that the models might heavily rely on specific urban traffic scenarios. Therefore, the performance of the models might vary in different types or scales of DTN networks, such as disaster recovery, military communication, or deep-sea exploration, among others. These scenarios possess unique network characteristics and requirements that could impact the accuracy of the model's predictions. In conclusion, while the models perform well in specific datasets and scenarios, their generalization capability and applicability in different types or scales of DTN networks need further validation.

### 5.4. Comparison to prior research

In this study, specific routing protocols and machine learning models were employed to predict optimal parameter combinations for enhancing delivery_prob while reducing overhead_ratio, thereby optimizing DTN energy consumption. However, compared to previous research in the field of DTN energy optimization, this approach has certain limitations.

Firstly, Research by Milena Radenkovic et al[16]. also focuses on using Machine Learning and San Francisco taxi mobility traces to optimize DTN Energy Consumption. Their proposed "CognitiveCharge" method is significantly unique in that it not only accurately identifies nodes and network areas at risk of energy depletion, but also dynamically reallocates energy from energy-abundant areas to those that are energy-poor. This strategy ensures optimal energy usage across the entire network. Its high accuracy and efficiency in DTN energy management are the directions that I need to study in my future research.

Secondly, The research by Vu San Ha Huynh et al[14]. on edge cloud services, represented by the adaptive ARPP protocol, employs real-time dynamic resource adjustments, utilizes distributed predictive analysis, and employs deep reinforcement learning to predict spatiotemporal patterns. They establish benchmarks for resource efficiency. Integrating these methodologies into our DTN study can facilitate precise energy allocation, balance operational efficiency, and enhance our energy consumption prediction methods across different network regions and time frames.

Additionally, the network topology was not fully considered, although it is a crucial factor affecting DTN energy consumption. PAOLO SANTI[51] pointed out that topology control (TC) is a vital approach in wireless ad hoc and sensor networks to reduce energy consumption and radio interference. TC can maintain the global connectivity properties of the network while minimizing energy consumption and interference related to links, providing a more comprehensive energy optimization.

Furthermore, this method did not introduce intermediate nodes like the "Throwbox" as in the work of Nilanjan Banerjee et al.[39] Throwbox, acting as fixed battery-powered nodes, can increase transmission opportunities in DTNs, thereby improving packet delivery rates and reducing message transmission delays. More importantly, the Throwbox architecture also pays special attention to power management to support long-term operation.

Finally, compared to research focusing on network dynamics, such as Lindgren et al.'s [52]PRoPHET algorithm, this model primarily relies on historical data for prediction. In highly dynamic and unpredictable network environments, this approach might limit the model's performance.

In conclusion, in order to further optimize DTN Energy Consumption, future work should consider CognitiveCharge, ARPP protocol, Network Topology, Network Dynamics, and possibly introduce intermediate nodes like Throwbox.

# VI. CONCLUSION

## 6.1. Main findings

The objective of this study is to optimize DTN energy consumption through the application of specific routing protocols and machine learning models. Specifically, the study utilizes the ONE simulator to conduct multiple simulations using external taxi data mobility traces from San Francisco. The goal is to predict optimal parameter combinations that enhance delivery_prob while reducing overhead_ratio. Among them, btInterface.transmitRange, btInterface.transmitSpeed, and Events1.size have a greater impact on the target variable. During the data analysis phase, two machine learning models, Random Forest and GBM, were employed, with a R-squared of 0.53 for Random Forest and 0.65 for GBM. These models predicted the values of delivery_prob and overhead_ratio through the differential differentiation method. The final outcomes demonstrate that the Random Forest model predicted a delivery_prob of 0.59 and overhead_ratio of 3.95, while the GBM model predicted a delivery_prob of 0.66 and overhead_ratio of 4.69. In comparison to the original simulations conducted with the ONE simulator, both models effectively succeeded in simultaneously increasing delivery_prob while decreasing overhead_ratio, thereby efficiently optimizing DTN energy consumption. Such results are closely related to the ability of machine learning models to capture and parse complex patterns and dependencies among various parameters.[53] This recognition capability allows the model to adopt more flexible and dynamic routing strategies, effectively improving the overall efficiency of the network. Especially in dealing with high-dimensional data and capturing complex nonlinear relationships, random forest and GBM models have shown strong performance.

## 6.2. Discussion

Firstly, the model exhibits a strong dependency on the dataset, primarily based on the taxi dataset from San Francisco for training and testing. Although it improves the realism of the experiment, it may limit the generalization ability of the model to other types of networks or geographical locations. Secondly, there is possibility for improvement in the model's R-squared value, indicating that both prediction accuracy and generalization capabilities could be enhanced.

Furthermore, although Random Forest and GBM models are capable of capturing high-dimensional and complex nonlinear relationships, they also sacrifice interpretability. This implies that it becomes challenging to precisely describe how altering a specific model parameter affects network performance.

In terms of scalability, this study mainly focuses on two performance indicators: delivery_prob and overhead_ratio. However, network performance may also be affected by many other factors, such as latency, throughput, reliability and error rate, etc., which limits the comprehensive optimization of DTN networks.

From a timeliness perspective, the model mainly relies on historical data for prediction and does not consider real-time changes in network status. This can be a problem in highly dynamic and unpredictable network environments.

From the overall perspective of the experiment, during the San Francisco taxi data conversion phase, we only introduced the movement trajectories of 40 taxis as data sources. This data scale may not be sufficient for overall analysis. Secondly, regarding the conversion of longitude and latitude coordinates to the coordinates of the simulator world, we use a linear mapping method. Although this method is simple and easy to implement, it may not be accurate enough. In terms of model construction, we observed that the R-squared of the model was not high enough, indicating that the model's prediction accuracy and generalization ability were insufficient. In terms of ONE simulator parameter combination prediction, we ran multiple times and obtained the best parameters.

Finally, due to the inherent complexity of DTN networks, apart from the considered parameters, there might exist other unconsidered but equally crucial parameters, such as Node Density, Packet Size, and Network Layer Protocols.

In summary, although this study achieved some optimization under specific conditions, these limitations also point out possible directions for future research to more comprehensively and accurately optimize the performance and energy efficiency of DTN networks.

## 6.3. Future work

Future research directions for further improving the energy consumption of DTN networks can be explored from multiple perspectives.

Firstly, the consideration of the Spray and Wait routing scheme, specifically designed for DTNs by Spyropoulos et al.[38], holds promise for optimizing energy usage. This scheme has demonstrated advantages over other routing methods, such as Epidemic routing, and could potentially enhance both delivery_prob and reduce overhead_ratio.

Secondly, to address the problem that the R-squared value of the current model is not high enough, future research can explore the use of more advanced machine learning algorithms, such as XGBoost, LightGBM or neural networks, and adjust model parameters to improve prediction accuracy and generalization capabilities.

From the perspective of the scalability of the model, future work can introduce more factors that affect network performance, such as latency, throughput, reliability, and error rate. Constructing models that aim to simultaneously improve throughput, reliability, and delivery_prob while decreasing latency, error rate, and overhead_ratio may be instrumental.

In addition to relying on simulation experiments, conducting tests in real-world network environments would be crucial for accurately assessing the model's effectiveness. Real-world testing would better capture the model's performance in the face of unforeseen factors like network congestion or hardware failures.

To enhance the model's generalization capability, the use of various types and sources of datasets for training and testing could be considered. For example, data from different types of transportation (such as buses, bicycles) or taxi data from diverse cities or regions could be employed. Even training and testing the model with entirely different types of network data could be explored.

In conclusion, future research directions are diverse, aiming to further optimize the performance and energy efficiency of DTN networks. In addition to the aspects mentioned above, emphasis should also be placed on network topology and dynamics. As noted by PAOLO SANTI[51], approaches like topology control (TC) can maintain global connectivity attributes of the network while minimizing energy consumption and link-related interference, providing a more comprehensive energy optimization. In addition, following the work of Nilanjan Banerjee et al.[39], it is also worthy of attention to consider the introduction of stationary battery-powered nodes like Throwbox. Such nodes can increase transmission opportunities in DTN and reduce data transmission delays, thereby further improving network performance and energy efficiency.

## VII. REFERENCES

[1] K. Fall and S. Farrell, "DTN: An architectural retrospective," IEEE Journal on Selected Areas in Communications, vol. 26, no. 5, 2008, doi: 10.1109/JSAC.2008.080609.

[2] X. Zhang, G. Neglia, J. Kurose, and D. Towsley, "Performance modeling of epidemic routing," Computer Networks, vol. 51, no. 10, 2007, doi: 10.1016/j.comnet.2006.11.028.

[3] J. Burgess, B. Gallagher, D. Jensen, and B. N. Levine, "MaxProp: Routing for vehicle-based disruption-tolerant networks," in Proceedings - IEEE INFOCOM, 2006. doi: 10.1109/INFOCOM.2006.228.

[4] M. Schonlau and R. Y. Zou, "The random forest algorithm for statistical learning," Stata Journal, vol. 20, no. 1, 2020, doi: 10.1177/1536867X20909688.

[5] A. Natekin and A. Knoll, "Gradient boosting machines, a tutorial," Front Neurorobot, vol. 7, no. DEC, 2013, doi: 10.3389/fnbot.2013.00021.

[6] P. R. Pereira, A. Casaca, J. J. P. C. Rodrigues, V. N. G. J. Soares, J. Triay, and C. Cervelló-Pastor, "From delay-tolerant networks to vehicular delay-tolerant networks," IEEE Communications Surveys and Tutorials, vol. 14, no. 4, 2012, doi: 10.1109/SURV.2011.081611.00102.

[7] X. Li, W. Shu, M. Li, H. Huang, and M. Y. Wu, "DTN routing in vehicular sensor networks," in GLOBECOM - IEEE Global Telecommunications Conference, 2008. doi: 10.1109/GLOCOM.2008.ECP.150.

[8] K. Fall, "A delay-tolerant network architecture for challenged internets," 2003. doi: 10.1145/863955.863960.

[9] M. Radenkovic and A. Grundy, "Efficient and adaptive congestion control for heterogeneous delay-tolerant networks," Ad Hoc Networks, vol. 10, no. 7, 2012, doi: 10.1016/j.adhoc.2012.03.013.

[10] S. Bhattacharjee, S. Roy, and S. Bandyopadhyay, "Exploring an energy-efficient DTN framework supporting disaster management services in post disaster relief operation," Wireless Networks, vol. 21, no. 3, 2015, doi: 10.1007/s11276-014-0836-5.

[11] S. L. F. Maia, É. R. Silva, and P. R. Guardieiro, "A new optimization strategy proposal for multi-copy forwarding in energy constrained DTNs," IEEE Communications Letters, vol. 18, no. 9, 2014, doi: 10.1109/LCOMM.2014.2346488.

[12] L. Junhai, Y. Danxia, X. Liu, F. Mingyu, and A. Voyiatzis, "A Survey of Delay-

and Disruption-Tolerant Networking Applications," Journal of Internet Engineering, vol. 5, no. 1, 2012.

[13] M. R. Schurgot, C. Comaniciu, and K. Jaffrès-Runser, "Beyond traditional DTN routing: Social networks for opportunistic communication," IEEE Communications Magazine, vol. 50, no. 7, 2012, doi: 10.1109/MCOM.2012.6231292.

[14] V. S. H. Huynh, M. Radenkovic, and N. Wang, "Distributed Spatial-Temporal Demand and Topology Aware Resource Provisioning for Edge Cloud Services," in 2021 6th International Conference on Fog and Mobile Edge Computing, FMEC 2021, 2021. doi: 10.1109/FMEC54266.2021.9732562.

[15] M. Radenkovic and V. S. Ha Huynh, "Energy-Aware Opportunistic Charging and Energy Distribution for Sustainable Vehicular Edge and Fog Networks," in 2020 5th International Conference on Fog and Mobile Edge Computing, FMEC 2020, 2020. doi: 10.1109/FMEC49853.2020.9144973.

[16] M. Radenkovic and A. Walker, "CognitiveCharge: Disconnection Tolerant Adaptive Collaborative and Predictive Vehicular Charging," Proceedings of the 4th ACM MobiHoc Workshop on Experiences with the Design and Implementation of Smart Objects, 2018.

[17] C. Lee, J. Lindh, and M. Hernes, "Measuring Bluetooth® Low Energy Power Consumption," Texas Instruments, no. 2017, 2015.

[18] F. M. Al-Turjman, A. E. Al-Fagih, W. M. Alsalih, and H. S. Hassanein, "A delay-tolerant framework for integrated RSNs in IoT," Comput Commun, vol. 36, no. 9, 2013, doi: 10.1016/j.comcom.2012.07.001.

[19] N. Benamar, K. D. Singh, M. Benamar, D. El Ouadghiri, and J. M. Bonnin, "Routing protocols in Vehicular Delay Tolerant Networks: A comprehensive survey," Comput Commun, vol. 48, 2014, doi: 10.1016/j.comcom.2014.03.024.

[20] M. C. Domingo, "An overview of the internet of underwater things," Journal of Network and Computer Applications, vol. 35, no. 6, 2012, doi: 10.1016/j.jnca.2012.07.012.

[21] C. S. De Abreu and R. M. Salles, "Modeling message diffusion in epidemical DTN," Ad Hoc Networks, vol. 16, 2014, doi: 10.1016/j.adhoc.2013.12.013.

[22] W. D. Ivancic, "Security analysis of DTN architecture and bundle protocol specification for space-based networks," in IEEE Aerospace Conference Proceedings, 2010. doi: 10.1109/AERO.2010.5446946.

[23] W. Zhao, M. Ammar, E. Zegura, and C. Computing, "A Message Ferrying Approach for Data Delivery in Sparse Mobile Ad Hoc Networks Categories and Subject Descriptors," MobiHoc, 2004.

[24] S. Sudevalayam and P. Kulkarni, "Energy harvesting sensor nodes: Survey and implications," IEEE Communications Surveys and Tutorials, vol. 13, no. 3, 2011, doi: 10.1109/SURV.2011.060710.00094.

[25] M. Demirbas and H. Ferhatosmanoglu, "Peer-to-peer spatial queries in sensor networks," in Proceedings - 3rd International Conference on Peer-to-Peer Computing, P2P 2003, 2003. doi: 10.1109/PTP.2003.1231501.

[26] Y. Zeng, K. Xiang, D. Li, and A. V. Vasilakos, "Directional routing and scheduling for green vehicular delay tolerant networks," Wireless Networks, vol. 19, no. 2, 2013, doi: 10.1007/s11276-012-0457-9.

[27] E. Altman, A. P. Azad, T. Başar, and F. De Pellegrini, "Optimal activation and transmission control in delay tolerant networks," in Proceedings - IEEE INFOCOM, 2010. doi: 10.1109/INFCOM.2010.5462264.

[28] H. Jun, M. H. Ammar, and E. W. Zegura, "Power management in delay tolerant networks: A framework and knowledge-based mechanisms," in 2005 Second Annual IEEE Communications Society Conference on Sensor and AdHoc Communications and Networks, SECON 2005, 2005. doi: 10.1109/SAHCN.2005.1557095.

[29] D. G. Reina, M. Askalani, S. L. Toral, F. Barrero, E. Asimakopoulou, and N. Bessis, "A Survey on Multihop Ad Hoc Networks for Disaster Response Scenarios," International Journal of Distributed Sensor Networks, vol. 2015. 2015. doi: 10.1155/2015/647037.

[30] X. Lu and P. Hui, "An energy-efficient n-epidemic routing protocol for Delay Tolerant Networks," in Proceedings - 2010 IEEE International Conference on Networking, Architecture and Storage, NAS 2010, 2010, pp. 341–347. doi: 10.1109/NAS.2010.46.

[31] T. Abdelkader, K. Naik, A. Nayak, N. Goel, and V. Srivastava, "A performance comparison of delay-tolerant network routing protocols," IEEE Netw, vol. 30, no. 2, 2016, doi: 10.1109/MNET.2016.7437024.

[32] Y. Wang, H. Dang, and H. Wu, "A survey on analytic studies of Delay-Tolerant mobile sensor networks," Wirel Commun Mob Comput, vol. 7, no. 10, 2007, doi: 10.1002/wcm.519.

[33] A. Jindal and K. Psounis, "Performance analysis of epidemic routing under contention," in IWCMC 2006 - Proceedings of the 2006 International Wireless Communications and Mobile Computing Conference, 2006. doi: 10.1145/1143549.1143657.

[34] A. Mukherjee, N. Dey, R. Kumar, B. K. Panigrahi, A. E. Hassanien, and J. M. R. S.

Tavares, "Delay Tolerant Network assisted flying Ad-Hoc network scenario: modeling and analytical perspective," Wireless Networks, vol. 25, no. 5, 2019, doi: 10.1007/s11276-019-01987-8.

[35] J. Philo, O. Hidekazu, W. Yong, M. Margaret, L. S. Peh, and D. Rubenstein, "Energy-efficient computing for wildlife tracking: Design tradeoffs and early experiences with ZebraNet," in Operating Systems Review (ACM), 2002. doi: 10.1145/635508.605408.

[36] R. C. Shah and J. M. Rabaey, "Energy aware routing for low energy ad hoc sensor networks," IEEE Wireless Communications and Networking Conference, WCNC, vol. 1, 2002, doi: 10.1109/WCNC.2002.993520.

[37] C. Schurgers and M. B. Srivastava, "Energy efficient routing in wireless sensor networks," in Proceedings - IEEE Military Communications Conference MILCOM, 2001. doi: 10.1109/milcom.2001.985819.

[38] T. Spyropoulos, K. Psounis, and C. S. Raghavendra, "Spray and wait: An efficient routing scheme for intermittently connected mobile networks," in Proceedings of the ACM SIGCOMM 2005 Workshop on Delay-Tolerant Networking, WDTN 2005, 2005. doi: 10.1145/1080139.1080143.

[39] N. Banerjee, M. D. Corner, and B. N. Levine, "An energy-efficient architecture for DTN throwboxes," in Proceedings - IEEE INFOCOM, 2007. doi: 10.1109/INFCOM.2007.96.

[40] W. Zhao, Y. Chen, M. Ammar, M. Corner, B. Levine, and E. Zegura, "Capacity enhancement using throwboxes in DTNs," in 2006 IEEE International Conference on Mobile Ad Hoc and Sensor Systems, MASS, 2006. doi: 10.1109/MOBHOC.2006.278570.

[41] J. Wu, Y. Zhu, L. Liu, B. Yu, and J. Pan, "Energy-efficient routing in multi-community DTN with social selfishness considerations," in 2016 IEEE Global Communications Conference, GLOBECOM 2016 - Proceedings, 2016. doi: 10.1109/GLOCOM.2016.7841809.

[42] B. Yang and X. Bai, "A Review of UAV Ferry Algorithms in Delay Tolerant Network," in Proceedings - 2019 12th International Symposium on Computational Intelligence and Design, ISCID 2019, 2019. doi: 10.1109/ISCID.2019.10104.

[43] G. Chen, B. Huang, X. Chen, L. Ge, M. Radenkovic, and Y. Ma, "Deep blue AI: A new bridge from data to knowledge for the ocean science," Deep-Sea Research Part I: Oceanographic Research Papers, vol. 190. 2022. doi: 10.1016/j.dsr.2022.103886.

[44] Y. Lu, W. Wang, L. Chen, Z. Zhang, and A. Huang, "Opportunistic forwarding in energy harvesting mobile delay tolerant networks," in 2014 IEEE International Conference on Communications, ICC 2014, 2014. doi: 10.1109/ICC.2014.6883372.

[45] B. Paramasivan, M. Bhuvaneswari, and K. M. Pitchai, "Augmented DTN based energy efficient routing protocol for vehicular ad hoc networks," in 2015 IEEE SENSORS - Proceedings, 2015. doi: 10.1109/ICSENS.2015.7370257.

[46] A. Keränen, "Opportunistic network environment simulator," Special Assignment report, Helsinki University of …, 2008.

[47] L. Breiman, "Random forests Machine Learning, 45 (1) (2001), pp. 5-32," Mach Learn, vol. 45, no. 1, 2001.

[48] S. B. Green, N. Redell, M. S. Thompson, and R. Levy, "Accuracy of Revised and Traditional Parallel Analyses for Assessing Dimensionality with Binary Data," Educ Psychol Meas, vol. 76, no. 1, 2016, doi: 10.1177/0013164415581898.

[49] F. Ghasemi, A. Mehridehnavi, A. Pérez-Garrido, and H. Pérez-Sánchez, "Neural network and deep-learning algorithms used in QSAR studies: merits and drawbacks," Drug Discovery Today, vol. 23, no. 10. 2018. doi: 10.1016/j.drudis.2018.06.016.

[50] D. Chicco, M. J. Warrens, and G. Jurman, "The coefficient of determination R-squared is more informative than SMAPE, MAE, MAPE, MSE and RMSE in regression analysis evaluation," PeerJ Comput Sci, vol. 7, 2021, doi: 10.7717/PEERJ-CS.623.

[51] P. Santi, "Topology control in wireless ad hoc and sensor networks," ACM Comput Surv, vol. 37, no. 2, 2005, doi: 10.1145/1089733.1089736.

[52] A. Lindgren, A. Doria, and O. Schelén, "Probabilistic routing in intermittently connected networks," Lecture Notes in Computer Science (including subseries Lecture Notes in Artificial Intelligence and Lecture Notes in Bioinformatics), vol. 3126, pp. 239–254, 2004, doi: 10.1007/978-3-540-27767-5_24.

[53] E. Christodoulou, J. Ma, G. S. Collins, E. W. Steyerberg, J. Y. Verbakel, and B. Van Calster, "A systematic review shows no performance benefit of machine learning over logistic regression for clinical prediction models," Journal of Clinical Epidemiology, vol. 110. 2019. doi: 10.1016/j.jclinepi.2019.02.004.

[54] Michal Piorkowski, Natasa Sarafijanovic-Djukic, Matthias Grossglauser, November 15, 2022, "CRAWDAD epfl/mobility", IEEE Dataport, doi: https://dx.doi.org/10.15783/C7J010.